\documentclass{article}

    \PassOptionsToPackage{numbers, compress}{natbib}

\usepackage[final]{neurips_2024}

\usepackage[utf8]{inputenc} 
\usepackage[T1]{fontenc}    
\usepackage{hyperref}       
\usepackage{url}            
\usepackage{booktabs}       
\usepackage{amsfonts}       
\usepackage{nicefrac}       
\usepackage{microtype}      
\usepackage[usenames, dvipsnames,table]{xcolor}  
\usepackage{multirow}       

\usepackage{caption}
\usepackage{subcaption}
\definecolor{Gray}{gray}{0.9}
\usepackage{wrapfig}

\usepackage{amsmath,amsfonts,bm}

\def\eqref#1{equation~\ref{#1}}

\def\1{\bm{1}}

\def\eps{{\epsilon}}

\DeclareMathAlphabet{\mathsfit}{\encodingdefault}{\sfdefault}{m}{sl}
\SetMathAlphabet{\mathsfit}{bold}{\encodingdefault}{\sfdefault}{bx}{n}

\DeclareMathOperator*{\argmax}{arg\,max}

\usepackage{amssymb} 

\usepackage{algorithm}
\usepackage{algpseudocode}

\algrenewcommand\algorithmicrequire{\textbf{Input:}}
\algrenewcommand\algorithmicensure{\textbf{Output:}}

\usepackage{enumitem}

\usepackage{bbm}

\usepackage{mathtools}

\usepackage{makecell}

\hypersetup{colorlinks,citecolor=NavyBlue,
linkcolor=NavyBlue,urlcolor=NavyBlue}
\usepackage[nameinlink,capitalise]{cleveref}

\newcommand{\ours}{{Genetic GFN}}

\title{Genetic-guided GFlowNets\\for Sample Efficient Molecular Optimization}

\author{%
  Hyeonah Kim$^1$\thanks{\texttt{hyeonah\_kim@kaist.ac.kr}} $\quad$ Minsu Kim$^1$ $\quad$ Sanghyeok Choi$^1$ $\quad$ Jinkyoo Park$^{1,2}$ \\
  $^1$Korea Advanced Institute of Science and Technology (KAIST), $\quad^2$OMELET\\
}

\begin{document}

\maketitle

\begin{abstract}
    

The challenge of discovering new molecules with desired properties is crucial in domains like drug discovery and material design. Recent advances in deep learning-based generative methods have shown promise but face the issue of sample efficiency due to the computational expense of evaluating the reward function.
This paper proposes a novel algorithm for sample-efficient molecular optimization by distilling a powerful genetic algorithm into deep generative policy using GFlowNets training, the off-policy method for amortized inference. This approach enables the deep generative policy to learn from domain knowledge, which has been explicitly integrated into the genetic algorithm. 
Our method achieves state-of-the-art performance in the official molecular optimization benchmark, significantly outperforming previous methods. It also demonstrates effectiveness in designing inhibitors against SARS-CoV-2 with substantially fewer reward calls.

\end{abstract}

\section{Introduction}
Discovering new molecules is one of the fundamental tasks in the chemical domain, with applications in drug discovery \citep{hughes2011principles} and material design \citep{hains2010molecular}. Particularly, \textit{de novo} molecular design focuses on generating novel molecules with desired properties from scratch. In this context, deep learning-based generative methods have emerged, showing promising results (e.g., \cite{olivecrona2017molecular, gomez2018automatic, jin2018jtvae, bengio2021flow, lee2023mood}). However, these methods still face a key challenge: the reward function is computationally expensive (e.g., assessing binding affinity through docking simulations), while the molecule space is combinatorially large.

Sample-efficient molecular optimization is thus crucial for discovering high-reward molecular structures with limited reward calls, especially for real-world applicability. The recently proposed benchmark, Practical Molecular Optimization (PMO)~\citep{gao2022sample}, has extensively assessed the sample efficiency of various algorithms, including reinforcement learning \cite{olivecrona2017molecular,zhou2019moldqn}, active learning \cite{jain2022biological}, variational autoencoders \cite{gomez2018automatic, jin2018jtvae}, generative flow networks (GFlowNets) \cite{bengio2021flow,zhu2024sample_moo}, and classical optimization methods like Bayesian optimization \cite{tripp2021fresh} and genetic algorithms \cite{jensen2019graph, nigam2021stoned}. 
Interestingly, the PMO benchmark has revealed a shift in algorithm rankings, with classical algorithms often outperforming recently proposed methods such as GFlowNets when the sample efficiency is considered.

Recent investigations \citep{Nigam2020AugmentingGA,tripp2023genetic}, including those highlighted by the PMO benchmark \cite{gao2022sample}, indicate that the classical frameworks, especially genetic algorithms (GA), still exhibit competitive performances compared to recently proposed deep learning methods.
These studies underscore that GAs effectively navigate the chemical space using domain-specific genetic operators. In contrast, deep learning methods usually do not leverage domain-specific knowledge, relying instead on deep networks to autonomously learn these insights. It can lead to inefficient training processes due to the lack of expert guidance \cite{ahn2020guiding}. To address this limitation, Genetic Expert Guided Learning (GEGL) \cite{ahn2020guiding} has been introduced, which enhances deep learning by distilling GA-generated samples into a deep generative policy using maximum likelihood estimation. This approach enables the deep generative policy to implicitly utilize domain-specific knowledge from the GA as a form of guidance. However, despite its successes, GEGL may face challenges in generalizing to unexplored regions and in learning a peaky distribution from samples, as it maximizes likelihood equally across all high-reward samples without adequately configuring the reward landscape.

\begin{figure*}
    \centering
    \vspace{-5pt}
    \includegraphics[width=0.94\linewidth]{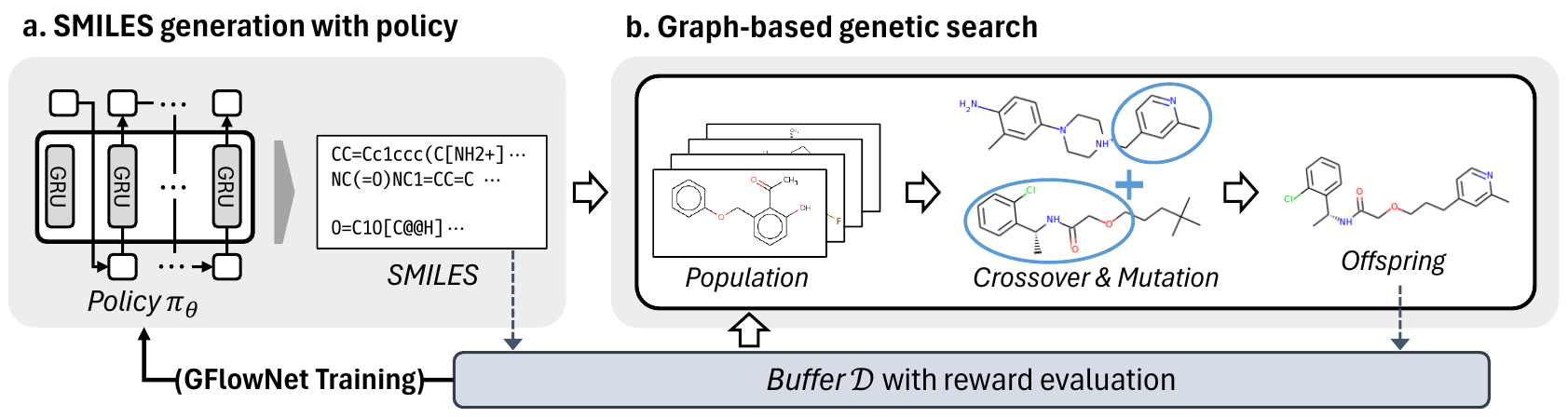}
    \caption{{Overview of \ours{}}. Our generative policy is trained to sample molecules proportional to rewards, and the genetic search refines them to higher-reward samples.}
    \label{fig:overview}
\end{figure*}

To address this, we propose a novel molecular optimization algorithm that integrates domain-specialized genetic algorithms into GFlowNets, an off-policy training method that trains policy to sample proportional to their rewards. 
As illustrated in \cref{fig:overview}, we first generate diverse candidates using the current policy and refine the candidates into higher-reward samples using GAs. Subsequently, we fine-tune the policy with a GFlowNet using the collected samples. Unlike the MLE training, which can be stuck in local modes when the reward landscape is peaky, GFlowNet trains the policy to sample molecules \emph{proportional to rewards}.
To enhance sample efficiency, we perform unsupervised pretraining on chemical datasets and regularize the GFlowNets policy with KL-divergence to align generative probability with the dataset distribution, focusing on the compact valid space.

Our contributions can be interpreted from both perspectives of GFlowNets and GAs:

\begin{list}{}{\setlength{\leftmargin}{1.2em}\setlength{\rightmargin}{0.5em}}

\item \textbf{GA increases the exploitation power of GFlowNets.} The proposed algorithm incorporates an effective off-policy exploration into GFlowNets based on domain-specialized GA. This approach aligns with recent studies that utilize off-policy explorations to guide toward the high-reward regions~\citep{kim2023lsgfn}. Our key contribution lies in explicitly leveraging domain knowledge about chemical structures and effectively distilling it into GFlowNets, which enable exceptional performance in real-world tasks beyond small-scale molecular generations. This contribution is crucial for the field of GFlowNets, which have struggled with sample-efficient molecular optimization tasks, even when using proxy reward models for active learning  \cite{jain2022biological,gao2022sample}; see \cref{sec:gfn}.

\item
\textbf{GFlowNets increases population diversity of GA.}
The proposed algorithm generates diversified samples using GFlowNets, enabling GA to effectively improve samples; our method can be regarded as a new genetic algorithm with deep generative policy-based population resetting. Our GFlowNet policy, parameterized over the whole space, periodically resets the population by diversely sampling individuals proportional to rewards, mitigating premature convergence of GA \citep{fogel1994introduction,pandey2014comparative}.
Experiments show that ours outperforms recent GAs in the PMO benchmark; see \cref{sec:exp_pmo}.
\end{list}

Our extensive experiments demonstrate the effectiveness and practical applicability of the proposed method.
First, our method achieves the highest total score of 16.213 across 23 oracles in the Practical Molecular Optimization benchmark \cite{gao2022sample}, outperforming all other baselines. 
Second, we conduct in \textit{in silico} experiments for designing SARS-CoV-2 inhibitors. The proposed method successfully generates inhibitors with ten times fewer reward calls. Moreover, our method effectively balances optimization and diversity, achieving higher scores with increased diversity compared to previously top-ranked methods.


\section{Background}
\subsection{Sample efficient \textit{de novo} molecular optimization}

Molecular design is the process of proposing new molecules likely to exhibit desirable outcomes. Compared to traditional virtual screening approaches, which identify suitable molecules from virtual libraries with a large number of molecules known a priori, de novo approaches seek to generate molecule structures anew.
The desired properties can be measured using score functions $\mathcal{O}$, called oracle. Formally, molecular design can be formulated as $\argmax_{x \in \mathcal{X}} \mathcal{O}(x)$,
where $x$ is a molecule, and $\mathcal{X}$ denotes the chemical space which comprises all possible molecules. 

The publication of standard benchmarks and datasets has facilitated the assessment of \textit{de novo} design methods (e.g., GuacaMol \cite{brown2019guacamol}, Therapeutics Data Commons (TDC) \cite{huang2021therapeutics}). 
The score functions are designed to consider various properties, such as the presence and absence of substructures, similarity, isomers, structural features, physicochemical properties, biological activity, and binding affinity (i.e., docking score). Notably, the PMO benchmark~\citep{gao2022sample} offers a unified framework that comprehensively evaluates the sample efficiency of a range of molecular design methods.

\subsection{Generative flow networks}

Generative flow networks (GFlowNets) \cite{bengio2021gflownet} are introduced as a new class of probabilistic models to sample a discrete compositional object $x \in \mathcal{X}$ from the target distribution, i.e., $P(x) \propto {e^{-\mathcal{E}(x)}}$. In general, direct sampling from the target distribution is challenging since the partition function $Z = \sum_{x\in\mathcal{X}}{e^{-\mathcal{E}(x)}}$ is intractable when the sample space is combinatorially large.
Hence, GFlowNets sample an object from an unnormalized distribution as a constructive generative process, where discrete \textit{actions} iteratively modify a \textit{state} — a partially constructed object. We define a trajectory as $\tau = (s_0, \ldots, s_T)$, where $s_T$ is a terminal state corresponding to a fully constructed object $x$.

A GFlowNet models flow $F$ of particles along a directed acyclic graph (DAG). The source and sink nodes of the DAG correspond to the initial state $s_0$ and terminal states $s_T$, respectively. 
The \textit{trajectory flow} $F(\tau)$ is defined as a flow through the trajectory $\tau$, and the \textit{state flow} $F(s)$ is defined as the sum of trajectory flows that include the state $s$, i.e., $F(s) = \sum_{s \in \tau} F(\tau)$. The \textit{edge flow} $F(s \rightarrow s')$ the sum of trajectory flows through the edge from state $s$ to $s'$, i.e., $F(s \rightarrow s') = \sum_{(s,s') \in \tau} F(\tau)$.

From the flow function \( F \), we derive two policy distributions. The \textit{forward policy} \( P_F(s'|s) \) is the probability of transitioning from state \( s \) to its child state \( s' \), defined as the edge flow \( F(s \rightarrow s') \) normalized by the state flow \( F(s) \), i.e., \( P_F(s'|s) = F(s \rightarrow s') / F(s) \). Similarly, the \textit{backward policy} \( P_B(s|s') \) is the probability of moving from state \( s' \) to its parent state \( s \), defined as \( P_B(s|s') = F(s \rightarrow s') / F(s') \). Utilizing these forward and backward policies, GFlowNets can derive an optimal sampler \( P(s_T) = \prod P_F(s_t|s_{t-1}) = R(s_T) / Z \) if balance conditions (e.g., \cite{bengio2021flow,malkin2022trajectory,madan2023sub-tb,pan2023fl-gfn,jang2024ledgfn}) are satisfied.

\paragraph{Trajectory balance loss \citep{malkin2022trajectory}.}
One of the most popular conditions is \textit{trajectory balance (TB)}, which directly parameterizes $P_F$, $P_B$, and flow of initial state (i.e., partition function) $Z$ to satisfy the following trajectory balance condition:
\begin{equation*}
    Z\prod_{t=1}^n P_F(s_t|s_{t-1}) = R(s_T) \prod_{t=1}^n P_B(s_{t-1}|s_t).
\end{equation*}
Then, this equation is converted into a loss function to be minimized along sampled trajectories, i.e., 
\begin{equation} \label{eq:tb}
    \mathcal{L}_{\text{TB}}(\tau; \theta) = \left( \log \frac{Z_\theta\prod_{t=1}^n P_F(s_t|s_{t-1};\theta)}{R(x)\prod_{t=1}^n P_B(s_{t-1}|s_t; \theta)} \right)^2.
\end{equation}

In GFlowNet training, employing exploratory behavior policies or replay training is allowed since GFlowNet can be trained in an off-policy manner, which is a key advantage \citep{bengio2021gflownet,malkin2022trajectory,zimmermann2023a,kim2023lsgfn}.

\section{Genetic-guided GFlowNets} \label{sec:method}

This section describes how the desired molecules are discovered with \ours{}. We model the generation process of molecules as a string-based constructive process. First, we pretrain the policy to learn the distribution of valid molecules. 
During the optimization phase, we iteratively generate molecules and update the policy with GFlowNet training using high-reward molecules. Particularly, we introduce graph-based genetic search to refine generated samples.

\subsection{Factorized string-based generative policy and unsupervised pretraining}
Building on insights from previous works~\citep{olivecrona2017molecular,ahn2020guiding}, we employ a string-based representation strategy, simplifying the molecular generation process by reducing it to a one-directional sequence generation. 
We adopt a sequence generative policy using a string-based assembly strategy, especially the simplified molecular-input line-entry system (SMILES) \cite{weininger1988smiles}. 
Motivated by REINVENT \citep{olivecrona2017molecular}, we parameterize the policy using a recurrent neural network architecture \cite{chung2014gru}.
Then, the probability $\pi_\theta(\bm{x})$ of generating a molecule, can be factorized to $\prod_{t=1}^{n} \pi_\theta(x_{t}|x_1, \ldots, x_{t-1})$, where $x_1, \ldots, x_n$ are characters of SMILES representation of $\bm{x}$. 

As demonstrated in previous studies, including \cite{olivecrona2017molecular, ahn2020guiding, gao2022sample}, pretraining is inevitable since training the generative policy from scratch is excessively sample-inefficient. Therefore, our policy is pre-trained to maximize the likelihood of valid molecules on existing chemical datasets $\mathcal{D}_{\text{pre}}$; note that pretraining \emph{does not require oracle information}. Precisely, the policy is pretrained to minimize the following:
\begin{equation}
    \mathcal{L}_{\text{pre}}(\bm{x}) = - \sum_{t=1}^n \log \pi_\theta(x_t|x_1, \ldots , x_{t-1}).
\end{equation}

\subsection{GFlowNet training of the generative policy with graph-based genetic search}


\begin{algorithm}[t]
   \caption{\ours{} training with limited reward calls} \label{alg:ours}
\begin{algorithmic}[1]
    \State Set $\pi_\theta \gets \pi_{\text{pre}}$, $\mathcal{D} \gets \emptyset$ 
    \While{$|\mathcal{D}| \leq \texttt{numOracle}$}
    \State $\mathcal{D} \gets \mathcal{D} \cup \{\bm{x}, \mathcal{O}(\bm{x}) \}$, where $\bm{x} \sim \pi_\theta (\cdot)$ \Comment{\textcolor{purple}{\textit{SMILES generation with policy}}}
    \State Initialize population $\mathcal{D}_{\text{pop}}$ from $\mathcal{D}$  \Comment{\textcolor{purple}{\textit{Graph-based genetic search}}}
    \For{$n=1$ {\bfseries to} \texttt{numGen}}
    \State $\bm{x} \gets \texttt{Crossover}(\bm{x}_{1}, \bm{x}_{2}),$ where $(\bm{x}_{1}, \bm{x}_{2}) \in \mathcal{D}_{\text{pop}}$
    \State $\bm{x}' \gets \texttt{Mutate}(\bm{x})$
    \State $\mathcal{D} \gets \mathcal{D} \cup \{\bm{x}', \mathcal{O}(\bm{x}') \}$, $\mathcal{D}_{\text{off}} \gets \mathcal{D}_{\text{off}} \cup \{\bm{x}', \mathcal{O}(\bm{x}') \}$
    \State $\mathcal{D}_{\text{pop}} \gets \texttt{Select}(\mathcal{D}_{\text{pop}} \cup \mathcal{D}_{\text{off}})$
    \EndFor
    \For{$k=1$ {\bfseries to} \texttt{numReplay}} \Comment{\textcolor{purple}{\textit{Updating the policy with GFlowNet training}}}
    \State Get $\mathcal{B}$ from $\mathcal{D}$ with rank-based sampling (\cref{eq:rank})
    \State Update $\theta$ to minimize $\frac{1}{|\mathcal{B}|}\sum_{\bm{x}\in \mathcal{B}}\mathcal{L}_{\text{TB}} + \alpha \text{KL}(\pi_\theta(\bm{x})||\pi_{\text{pre}}(\bm{x}))$ 
    \EndFor
    \EndWhile
\end{algorithmic}
\end{algorithm}

To generate desirable molecules with limited reward calls, we iteratively generate samples using two distinct strategies (\cref{sec:exploration}) and fine-tune the policy, initialized with $\pi_{\text{pre}}$, using a GFlowNet by replaying collected samples (\cref{sec:training}). The overall procedure is described in \cref{alg:ours}.

\subsubsection{Molecule generation strategies in \ours{}}
\label{sec:exploration}

We employ two distinct molecule generation strategies, SMILES generation with our training policy and graph-based genetic search. 
These two strategies are synergized to generate diversified and high-reward samples, efficiently searching the vast chemical space.

\textbf{SMILES generation with policy.} The training policy $\pi_\theta$ generates SMILES sequences. Since the policy is trained using the trajectory balance loss (see the following subsection), it is the same as sampling $\bm{x}$ from $\prod_{t=1}^T P_F(s_t|s_{t-1}) \propto R(s_T=\bm{x})$, where $s_{t-1}$ is represented by previously collected SMILES token, and $\log R(\bm{x})=-\beta \mathcal{O}(\bm{x})$ with an inverse temperature $\beta$.


\textbf{Graph-based genetic search.} To effectively search the higher-reward region, we employ a genetic algorithm that iteratively evolves populations through \textit{crossover}, \textit{mutation}, and \textit{selection}. 
We adopt the operations of the graph-based genetic algorithm, Graph GA \citep{jensen2019graph}, which has proven to effectively search the molecule space with finely designed genetic operations; please refer to the original paper for details. The genetic search is performed as follows:
\begin{enumerate}
    \item \textbf{Initialize a population} $\mathcal{D}_{\text{pop}}$: The initial population is selected from the whole buffer $\mathcal{D}$, consisting of samples from the policy and previous genetic search.
    \item \textbf{Generate offspring} $\mathcal{D}_{\text{off}}$: A child is generated from randomly chosen two parent molecules by combining the fragments (\textit{crossover}). Then, the child is randomly modified (\textit{mutation}). 
    \item \textbf{Select a new population} $\mathcal{D}'_{\text{pop}}$: Sample from $\mathcal{D}_{\text{pop}} \cup \mathcal{D}_{\text{off}}$, and go back to 2.
\end{enumerate}

One key advantage is that offspring can have a large distance from the parents in the 1D string space, even if the molecule distances are small, which is beneficial to avoid being stuck in local optima; see the experimental results in \cref{tab:dist}.




\subsubsection{Updating the generative policy with GFlowNets training} \label{sec:training}

Using the generated samples, the policy is fine-tuned using the trajectory balance loss. The off-policy property of GFlowNet losses enables the utilization of refined samples from the genetic search. In particular, for better sample efficiency, we employ replay training with a rank-based reweighed buffer.

\paragraph{TB loss with KL-divergence penalty.} 
The generative policy is trained using the trajectory balance loss in \cref{eq:tb}. 
Note that we set $P_B$ to 1 for simplicity since SMILES generations are conducted in one direction.
To ensure that the policy does not deviate excessively from the pretrained policy during training, we introduce a Kullback-Leibler (KL) divergence penalty inspired by the works in language model fine-tuning \cite{ziegler2019rlhf,rafailov2023dpo}. 
Thus, our model is updated to minimize the following loss function:
\begin{equation}
    \mathcal{L} = \mathcal{L}_{\text{TB}}(\tau; \theta) + \alpha \text{KL}(\pi_\theta(x)|| \pi_{\text{pre}}(x)),
\end{equation}
where $\pi_{\text{pre}}$ denotes the the pre-trained policy. As a result, $\pi_\theta$ is trained to generate desired (by $\mathcal{L}_{\text{TB}}$) and valid (by $\pi_{\text{pre}}$) molecules. Note that trajectories on which the proposed loss is minimized are sampled from the experience buffer.

\paragraph{Rank-based reweighed experience buffer.} 
The rank-based reweighting biases the samples towards high-reward candidates by assigning greater weight to trajectories with higher ranks, thereby enhancing the focus on more promising solutions \citep{tripp2020rank, kim2024bootgen}. The weight is computed as follows:
\begin{equation} \label{eq:rank}
    \frac{\left(k |\mathcal{D}| + \text{rank}_{\mathcal{O}, \mathcal{D}}(\bm{x})\right)^{-1}}{\sum_{\bm{x} \in \mathcal{D}} \left(k |\mathcal{D}| + \text{rank}_{\mathcal{O}, \mathcal{D}}(\bm{x})\right)^{-1}}.
\end{equation}
Here, $k$ is a weight-shifting factor, and $\text{rank}_{\mathcal{O}, \mathcal{D}}(\bm{x})$ is a relative rank of value of $\mathcal{O}(\bm{x})$ in the dataset $\mathcal{D}$. Note that we also utilize rank-based sampling in the genetic search (steps 1 and 3).

\section{Related works}

\subsection{Genetic algorithms for molecular optimization} \label{sec:ga}

Genetic algorithm (GA) is a representative meta-heuristic inspired by the evolutionary process. 
This subsection focuses on discussing the application of GA in molecular optimization. 
As one of the seminal works, a graph-based GA (Graph GA) was proposed with sophisticatedly designed operations based on chemical knowledge \cite{jensen2019graph}. 
Note that our method also adopts Graph GA operations in the genetic search.
Various strategies for molecular assembly, not limited to graphs, have been utilized in GA \citep{yoshikawa2018population,nigam2021stoned,gao2022synnet}. 
A recent contribution by \cite{tripp2023genetic} introduces an enhanced version of Graph GA. They introduce quantile-uniform sampling to bias the population towards containing higher reward samples while maintaining diversity. Experimental results from Mol GA demonstrate the effectiveness of GAs as strong baselines, achieving state-of-the-art (SOTA) performance in the PMO benchmark.

\subsection{GFlowNets for molecular optimization} \label{sec:gfn}

Generative Flow Networks (GFlowNets or GFN) \citep{bengio2021gflownet, bengio2021flow} have drawn significant attention in scientific discovery \citep{jain2023gflownets}, particularly in molecular optimization and biological sequence design \citep{bengio2021flow, jain2022biological, shen23guided_tb, kim2023lsgfn, ghari2023generative, cretu2024synflownet, zhu2024sample_moo,kim24srt}. GFlowNets, which are off-policy variational inference methods \cite{malkin2022gflownets}, are closely related to value-based reinforcement learning within soft Markov Decision Processes (soft MDPs) \citep{mohammadpour2023maximum, deleu2024discrete}, focusing on learning maximum entropy agents \citep{nachum2017bridging, haarnoja2017reinforcement}. This allows GFlowNets to generate diverse candidates for chemical and biological structures. Advances in GFlowNets have included new objective functions \citep{bengio2021flow, malkin2022trajectory, madan2023sub-tb}, improved credit assignments \citep{pan2023fl-gfn, jang2024ledgfn}, and enhanced off-policy exploration strategies \citep{kim2023lsgfn, rector2023thompson,  kim2023learning,guo2024dynamic}. Our approach, an improved off-policy exploration technique for GFlowNets, enhances sample efficiency and scalability for moderate-scale chemical discovery. We present extensive experiments targeting the discovery of larger-scale molecules, with SMILES strings approximately 100 characters long, surpassing the scope of smaller molecule generation tasks commonly addressed in GFlowNets literature \citep{bengio2021flow, shen23guided_tb}.

\section{Experiments}
This section provides extensive experimental results, including experiments on the official sample-efficient molecular optimization benchmark and \textit{in silico} design for SARS-CoV-2 inhibitors. The codes are available at \href{https://github.com/hyeonahkimm/genetic_gfn}{https://github.com/hyeonahkimm/genetic\_gfn}.

\begin{table*}
    \centering
    \caption{Mean and standard deviation of AUC top-10 ($\uparrow$) from five independent runs. We use oracle ID (in lexicographical order) instead of a name for better readability, and the best mean scores are denoted in \textbf{bold} for each task.
    The results of further baselines are provided in \cref{appnd:full_pmo}.}
    \resizebox{\textwidth}{!}{
    \begin{tabular}{lccccccc}
\toprule
ID & \makecell{\ours{}\\(Ours)} & Mol GA \cite{tripp2023genetic} & \makecell{SMILES\\ {REINVENT} \cite{olivecrona2017molecular}} & GEGL \cite{ahn2020guiding} & GP BO \cite{tripp2021fresh}  & \makecell{Fragment\\GFN \cite{bengio2021flow}} & \makecell{Fragment\\GFN-AL \cite{jain2022biological}}\\
\midrule
\#1 & \textbf{0.949 \footnotesize{$\pm$  0.010}} & 0.928 \footnotesize{$\pm$  0.015} & 0.881 \footnotesize{$\pm$  0.016} & 0.842 \footnotesize{$\pm$  0.019} & 0.902 \footnotesize{$\pm$  0.011} & 0.382 \footnotesize{$\pm$  0.010} & 0.459 \footnotesize{$\pm$  0.028} \\
\#2 & \textbf{0.761 \footnotesize{$\pm$  0.019}} & 0.740 \footnotesize{$\pm$  0.055} & 0.644 \footnotesize{$\pm$  0.019} & 0.626 \footnotesize{$\pm$  0.018} & 0.579 \footnotesize{$\pm$  0.035} & 0.428 \footnotesize{$\pm$  0.002} & 0.437 \footnotesize{$\pm$  0.007} \\
\#3 & \textbf{0.802 \footnotesize{$\pm$  0.029}} & 0.629 \footnotesize{$\pm$  0.062} & 0.717 \footnotesize{$\pm$  0.027} & 0.699 \footnotesize{$\pm$  0.041} & 0.746 \footnotesize{$\pm$  0.025} & 0.263 \footnotesize{$\pm$  0.009} & 0.326 \footnotesize{$\pm$  0.008} \\
\#4 & 0.733 \footnotesize{$\pm$  0.109} & 0.656 \footnotesize{$\pm$  0.013} & 0.662 \footnotesize{$\pm$  0.044} & 0.656 \footnotesize{$\pm$  0.039} & 0.615 \footnotesize{$\pm$  0.009} & 0.582 \footnotesize{$\pm$  0.001} & 0.587 \footnotesize{$\pm$  0.002} \\
\#5 & \textbf{0.974 \footnotesize{$\pm$  0.006}} & 0.950 \footnotesize{$\pm$  0.004} & 0.957 \footnotesize{$\pm$  0.007} & 0.898 \footnotesize{$\pm$  0.015} & 0.941 \footnotesize{$\pm$  0.017} & 0.480 \footnotesize{$\pm$  0.075} & 0.601 \footnotesize{$\pm$  0.055} \\
\#6 & \textbf{0.856 \footnotesize{$\pm$  0.039}} & 0.835 \footnotesize{$\pm$  0.012} & 0.781 \footnotesize{$\pm$  0.013} & 0.769 \footnotesize{$\pm$  0.009} & 0.726 \footnotesize{$\pm$  0.004} & 0.689 \footnotesize{$\pm$  0.003} & 0.700 \footnotesize{$\pm$  0.005} \\
\#7 & 0.881 \footnotesize{$\pm$  0.042} & \textbf{0.894 \footnotesize{$\pm$  0.025}} & 0.885 \footnotesize{$\pm$  0.031} & 0.816 \footnotesize{$\pm$  0.027} & 0.861 \footnotesize{$\pm$  0.027} & 0.589 \footnotesize{$\pm$  0.009} & 0.666 \footnotesize{$\pm$  0.006} \\
\#8 & \textbf{0.969 \footnotesize{$\pm$  0.003}} & 0.926 \footnotesize{$\pm$  0.014} & 0.942 \footnotesize{$\pm$  0.012} & 0.930 \footnotesize{$\pm$  0.011} & 0.883 \footnotesize{$\pm$  0.040} & 0.791 \footnotesize{$\pm$  0.024} & 0.468 \footnotesize{$\pm$  0.211} \\
\#9 & \textbf{0.897 \footnotesize{$\pm$  0.007}} & 0.894 \footnotesize{$\pm$  0.005} & 0.838 \footnotesize{$\pm$  0.030} & 0.808 \footnotesize{$\pm$  0.007} & 0.805 \footnotesize{$\pm$  0.007} & 0.576 \footnotesize{$\pm$  0.021} & 0.199 \footnotesize{$\pm$  0.199} \\
\#10 & 0.764 \footnotesize{$\pm$  0.069} & \textbf{0.835 \footnotesize{$\pm$  0.040}} & 0.782 \footnotesize{$\pm$  0.029} & 0.580 \footnotesize{$\pm$  0.086} & 0.611 \footnotesize{$\pm$  0.080} & 0.359 \footnotesize{$\pm$  0.009} & 0.442 \footnotesize{$\pm$  0.017} \\
\#11 & \textbf{0.379 \footnotesize{$\pm$  0.010}} & 0.329 \footnotesize{$\pm$  0.006} & 0.363 \footnotesize{$\pm$  0.011} & 0.338 \footnotesize{$\pm$  0.016} & 0.298 \footnotesize{$\pm$  0.016} & 0.192 \footnotesize{$\pm$  0.003} & 0.207 \footnotesize{$\pm$  0.003} \\
\#12 & 0.294 \footnotesize{$\pm$  0.007} & 0.284 \footnotesize{$\pm$  0.035} & 0.281 \footnotesize{$\pm$  0.002} & 0.274 \footnotesize{$\pm$  0.007} & \textbf{0.296 \footnotesize{$\pm$  0.011}} & 0.174 \footnotesize{$\pm$  0.002} & 0.181 \footnotesize{$\pm$  0.002} \\
\#13 & 0.708 \footnotesize{$\pm$  0.057} & \textbf{0.762 \footnotesize{$\pm$  0.048}} & 0.634 \footnotesize{$\pm$  0.042} & 0.599 \footnotesize{$\pm$  0.035} & 0.631 \footnotesize{$\pm$  0.093} & 0.291 \footnotesize{$\pm$  0.005} & 0.332 \footnotesize{$\pm$  0.012} \\
\#14 & \textbf{0.860 \footnotesize{$\pm$  0.008}} & 0.853 \footnotesize{$\pm$  0.005} & 0.834 \footnotesize{$\pm$  0.010} & 0.832 \footnotesize{$\pm$  0.005} & 0.788 \footnotesize{$\pm$  0.005} & 0.787 \footnotesize{$\pm$  0.002} & 0.785 \footnotesize{$\pm$  0.003} \\
\#15 & 0.595 \footnotesize{$\pm$  0.014} & \textbf{0.610 \footnotesize{$\pm$  0.038}} & 0.535 \footnotesize{$\pm$  0.015} & 0.537 \footnotesize{$\pm$  0.015} & 0.494 \footnotesize{$\pm$  0.006} & 0.423 \footnotesize{$\pm$  0.006} & 0.434 \footnotesize{$\pm$  0.006} \\
\#16 & \textbf{0.942 \footnotesize{$\pm$  0.000}} & 0.941 \footnotesize{$\pm$  0.001} & 0.941 \footnotesize{$\pm$  0.000} & 0.941 \footnotesize{$\pm$  0.001} & 0.937 \footnotesize{$\pm$  0.002} & 0.904 \footnotesize{$\pm$  0.002} & 0.917 \footnotesize{$\pm$  0.002} \\
\#17 & 0.819 \footnotesize{$\pm$  0.018} & \textbf{0.830 \footnotesize{$\pm$  0.010}} & 0.770 \footnotesize{$\pm$  0.005} & 0.730 \footnotesize{$\pm$  0.011} & 0.741 \footnotesize{$\pm$  0.010} & 0.626 \footnotesize{$\pm$  0.005} & 0.660 \footnotesize{$\pm$  0.004} \\
\#18 & \textbf{0.615 \footnotesize{$\pm$  0.100}} & 0.568 \footnotesize{$\pm$  0.017} & 0.551 \footnotesize{$\pm$  0.024} & 0.531 \footnotesize{$\pm$  0.010} & 0.535 \footnotesize{$\pm$  0.007} & 0.461 \footnotesize{$\pm$  0.002} & 0.464 \footnotesize{$\pm$  0.003} \\
\#19 & 0.634 \footnotesize{$\pm$  0.039} & \textbf{0.677 \footnotesize{$\pm$  0.055}} & 0.470 \footnotesize{$\pm$  0.041} & 0.402 \footnotesize{$\pm$  0.024} & 0.461 \footnotesize{$\pm$  0.057} & 0.180 \footnotesize{$\pm$  0.012} & 0.217 \footnotesize{$\pm$  0.022} \\
\#20 & \textbf{0.583 \footnotesize{$\pm$  0.034}} & 0.544 \footnotesize{$\pm$  0.067} & 0.544 \footnotesize{$\pm$  0.026} & 0.515 \footnotesize{$\pm$  0.028} & 0.544 \footnotesize{$\pm$  0.038} & 0.261 \footnotesize{$\pm$  0.004} & 0.292 \footnotesize{$\pm$  0.009} \\
\#21 & \textbf{0.511 \footnotesize{$\pm$  0.054}} & 0.487 \footnotesize{$\pm$  0.024} & 0.458 \footnotesize{$\pm$  0.018} & 0.420 \footnotesize{$\pm$  0.031} & 0.404 \footnotesize{$\pm$  0.025} & 0.183 \footnotesize{$\pm$  0.001} & 0.190 \footnotesize{$\pm$  0.002} \\
\#22 & 0.135 \footnotesize{$\pm$  0.271} & 0.000 \footnotesize{$\pm$  0.000} & \textbf{0.182 \footnotesize{$\pm$  0.363}} & 0.119 \footnotesize{$\pm$  0.238} & 0.000 \footnotesize{$\pm$  0.000} & 0.000 \footnotesize{$\pm$  0.000} & 0.000 \footnotesize{$\pm$  0.000} \\
\#23 & \textbf{0.552 \footnotesize{$\pm$  0.033}} & 0.514 \footnotesize{$\pm$  0.033} & 0.533 \footnotesize{$\pm$  0.009} & 0.492 \footnotesize{$\pm$  0.021} & 0.466 \footnotesize{$\pm$  0.025} & 0.308 \footnotesize{$\pm$  0.027} & 0.353 \footnotesize{$\pm$  0.024} \\
\midrule
Sum &  \textbf{16.213} & 15.686 & 15.185 & 14.354 & 14.264 & 9.929 & 9.917 \\
\bottomrule
\end{tabular}
}
    \label{tab:auc10}
\end{table*}

\begin{figure}
    \centering
    \includegraphics[width=\textwidth]{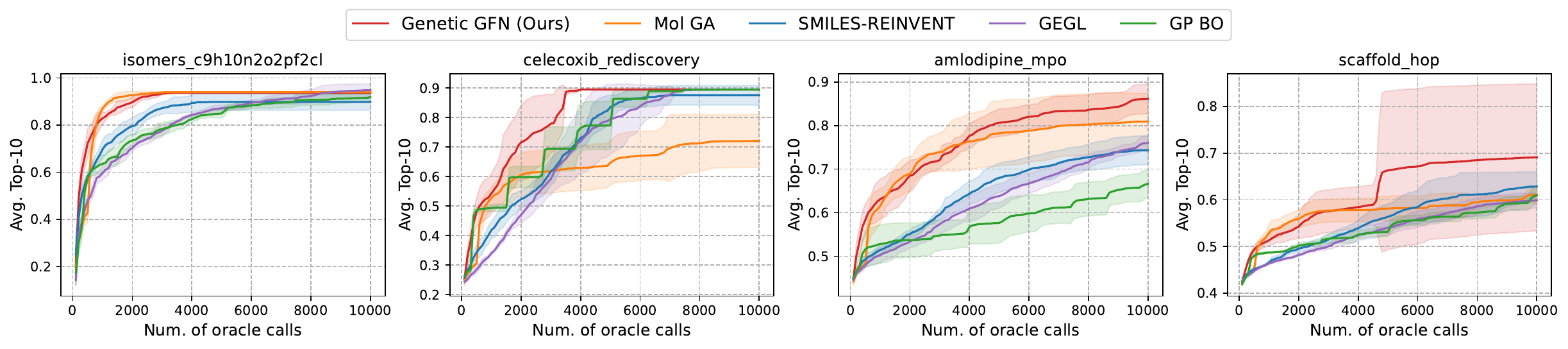}
    \caption{The optimization curve of the average scores of Top-10 over the score function calls. All optimization curves for 23 oracles are provided in \cref{appnd:full_pmo}.} 
    \label{fig:opt_curve}
\end{figure}

\subsection{Sample efficient molecular optimization} \label{sec:exp_pmo}

In this experiment section, we compare \ours{} with various molecular optimization methods from the perspective of sample efficiency. The performance is primarily quantified by the area under the curve (AUC), with the number of score function calls limited to 10K. Note that we rigorously follow the Practical Molecular Optimization (PMO) benchmark \cite{gao2022sample}.

\subsubsection{Main results in the official benchmark of PMO}

As baselines, we employ Top-8 methods from the PMO benchmark since they recorded the best AUC Top-10 in at least one oracle. The baseline methods include various ranges of algorithms and representation strategies. 
First, REINVENT \cite{olivecrona2017molecular} is an RL method that tunes the policy with adjusted likelihood.
Graph GA \cite{jensen2019graph}, STONED \cite{nigam2021stoned}, SMILES GA \cite{brown2019guacamol}, and SynNet \cite{gao2022synnet} are genetic algorithms that utilize different assembly strategies; they use fragment-based graphs, SELFIES, SMILES, and synthesis, respectively. Additionally, a hill climbing method (SMILES-LSTM-HC \cite{brown2019guacamol}) and Bayesian optimization (GP BO \cite{tripp2021fresh}) are included. SMILES-LSTM-HC iteratively generates samples and imitates high-reward samples, while GP BO uses a surrogate model with the Gaussian process (GP) and Graph GA to optimize the GP acquisition functions in the inner loop.

Moreover, we adopt additional methods, Mol GA \cite{tripp2023genetic} and GEGL \cite{ahn2020guiding}. Mol GA is an advanced version of Graph GA and outperforms other baselines in the PMO benchmark. On the other hand, GEGL is an ablated version of our approach that utilizes imitation learning with a reward-priority queue instead of GFlowNet training with rank-based sampling. For both, we adopt the original implementations\footnote{\href{https://github.com/AustinT/mol_ga}{https://github.com/AustinT/mol\_ga}}\footnote{\href{https://github.com/sungsoo-ahn/genetic-expert-guided-learning}{https://github.com/sungsoo-ahn/genetic-expert-guided-learning}} with hyperparameters searches following the guidelines; see \cref{appnd:tuning}.

The main results in \cref{tab:auc10} report the AUC score of Top-10 candidates with independent five runs with different seeds. In addition, \cref{fig:opt_curve} visually presents the Top-10 average score across the computational budget, i.e., the number of oracle calls, providing a concise overview of the results. Due to the lack of space, the best five results are provided; please check \cref{appnd:full_pmo} for the rest of the results.
As shown in \cref{tab:auc10}, \ours{} outperforms the other baselines with a total of 16.213 and attains the highest AUC Top-10 values in 14 out of 23 score functions. 
The results of diversity and SA score for each oracle are presented in \cref{appnd:div_sa}.

\subsubsection{Controllability of the scores-diversity trade-off and ablation studies} \label{sec:tradeoff_exp}

\begin{wrapfigure}{r}{0.47\linewidth}
    \centering
    \vspace{-8pt}
    \includegraphics[width=\linewidth]{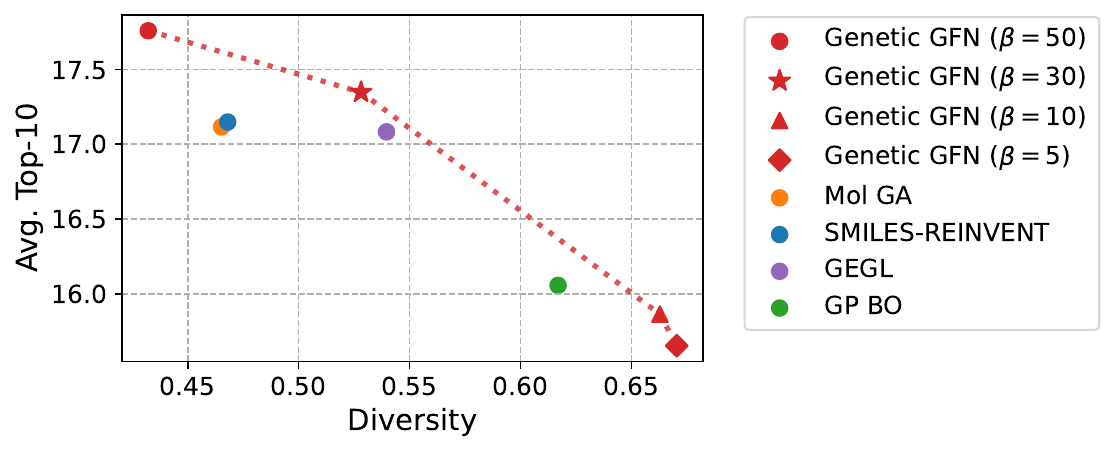}
    \resizebox{0.95\linewidth}{!}{\begin{tabular}{lcccccccc}
\toprule
 & $\beta = 1$ & $\beta = 5$ & $\beta = 10$ & $\beta = 30$ &  $\beta = 50$   \\
\midrule
\textbf{AUC-10} & 11.083 & 14.597 & 14.735 & 15.815 & \textit{16.213} \\
\textbf{Div.} & 0.812 & 0.670 & 0.663 & 0.528 & \textit{0.432} \\
\bottomrule
\end{tabular} 
}
    \caption{Average of Top-10 score and diversity. Note that the fragment-based GFlowNet achieves 10.957 with a diversity of 0.816.} \label{fig:beta_ablation}
    \vspace{-5pt}
\end{wrapfigure}
\paragraph{Controllability of the scores-diversity trade-off.} 
In the benchmark, we found a pronounced trade-off between attaining high evaluation scores within a limited budget and generating diverse molecular candidates. This section demonstrates the controllability of the score-diversity trade-off through adjustments in the inverse temperature $\beta$. 
Decreasing the inverse temperature gives more diverse candidates. 
The results in \cref{fig:beta_ablation} demonstrate adjustments of $\beta$ can control the trade-off between score and diversity, achieving Pareto-frontier to other baselines in the benchmark. 
Notably, \ours{} with $\beta=30$ achieves a higher AUC Top-10 with a greater diversity compared to the SOTA GA method (Mol GA: 15.686 with a diversity of 0.465) and RL method (REINVENT: 15.185 with a diversity of 0.468). 
Similarly, the weight-shifting factor $k$ in rank-based sampling can control the trade-off; see \cref{appnd:k}.

\begin{table}
\centering
\caption{Ablation studies. In the GS ablation study (-$\{\text{GS}\}$), the generative policy solely generates samples, while $\eps$-greedy samples from $P_F$ mixed with a uniform distribution. The \textbf{bold} text indicates the best value.} \label{tab:abl}
\resizebox{0.8\linewidth}{!}{\begin{tabular}{lcccc}
\toprule
 & \multirow{2}{*}{\ours{}} & \multicolumn{2}{c}{Genetic Search} & \multirow{2}{*}{\makecell{KL-divergence\\penalty}} \\
&  & - $\{\text{GS}\}$ & - $\{\text{GS}\}$ + $\{ \eps\text{-greedy} \}$ & \\
\midrule
AUC Top-1
& \textbf{16.530 \footnotesize{$\pm$ 0.198} }
& 16.070 \footnotesize{$\pm$ 0.290}
& 15.966 \footnotesize{$\pm$ 0.085}
& 16.251 \footnotesize{$\pm$ 0.440} \\
AUC Top-10 
& \textbf{16.213 \footnotesize{$\pm$ 0.173}}
& 15.738 \footnotesize{$\pm$ 0.274}
& 15.626 \footnotesize{$\pm$ 0.082}
& 15.928 \footnotesize{$\pm$ 0.426} \\
AUC Top-100
& \textbf{15.516 \footnotesize{$\pm$ 0.127}}  
& 15.030 \footnotesize{$\pm$ 0.322}
& 14.939 \footnotesize{$\pm$ 0.147}
& 15.188 \footnotesize{$\pm$ 0.297} \\
\bottomrule
\end{tabular}

}
\vspace{-7pt}
\end{table}

\textbf{Ablation studies.} 
The ablation studies investigate the essential components of our framework: the genetic search (GS) and the KL-divergence penalty.
To assess the effectiveness of the genetic search, we also compare its performance against the exploration strategy used in previous GFlowNet studies \cite{bengio2021flow, jain2022biological}. It samples actions from a GFlowNet sampler mixed with a uniform distribution, similar to $\epsilon$-greedy in RL.
The results, shown in \cref{tab:abl}, reveal that the removal of either component results in a decline in performance, underscoring the importance of employing a suitable exploration strategy. Detailed results, including statistical analysis, are provided in \cref{appnd:full_abl}.

\subsubsection{Comparisons with GFlowNets variants} \label{sec:exp_gfn}

We compare \ours{} with the graph-based GFlowNet \citep{bengio2021flow}, GFlowNet-AL \citep{jain2022biological}, and the local search GFlowNet (LS-GFN) \cite{kim2023lsgfn} using SMILES representations. LS-GFN utilizes Monte Carlo Markov Chain (MCMC) techniques, incorporating partial backtracking and reconstructing solution trajectories with the training policy as the proposal distribution \cite{kim2023lsgfn}.
The experiments are conducted on the PMO benchmark, and we implement LS-GFN with SMILES by replacing our genetic search with a local search. 
Note that while the original LS-GFN employs the prepend-append MDP, which does not directly apply to SMILES, we use the same one-directional SMILES generation as ours.

As shown in \cref{tab:gfn}, \ours{} outperforms other GFlowNet variants. Notably, generating SMILES is significantly more advantageous than generating graph-based fragments. The performance gap between \ours{} and LS-GFN highlights the importance of a proper exploratory policy. 
To further analyze, we measure the distance between samples before and after searches in GSK3$\beta$ and JNK3. The normalized Levenshtein distances for SMILES and Tanimoto similarity for molecules are reported in \cref{tab:dist}.
The results show that the local search may be inefficient in effectively searching in moderate-scale chemical spaces because its capabilities heavily depend on the current policy, leading to suboptimal search performance. In contrast, our approach leverages a domain-specialized genetic search within the molecule graph space, working as an effective off-policy exploration --- the samples with SMILES representation are used to train the string-based generative policy.

\begin{table}
\vspace{-10pt}
\caption{Comparison with GFlowNet variants. Notably, samples from the GS have larger SMILES distances than LS, leading to better sample efficiency. The \textbf{bold} text indicates the best value.}
\begin{subtable}{0.72\textwidth}
\centering
\caption{Average and standard deviation of AUC scores ($\uparrow$)}\label{tab:gfn}
\resizebox{\linewidth}{!}{\begin{tabular}{lcccc}
\toprule
 & \multicolumn{2}{c}{SMILES} & \multicolumn{2}{c}{Fragment-based} \\
\cmidrule(r){2-3} \cmidrule(r){4-5}
 &  \ours{} & LS-GFN \cite{kim2023lsgfn}  & GFN \cite{bengio2021flow} & GFN-AL \cite{jain2022biological} \\
\midrule
{AUC Top-1} 
& \textbf{16.530 \footnotesize{$\pm$ 0.198}}
& 15.514 \footnotesize{$\pm$ 0.269}
& 10.957 \footnotesize{$\pm$ 0.033}
& 11.032 \footnotesize{$\pm$ 0.016} \\
{AUC Top-10} 
& \textbf{16.213 \footnotesize{$\pm$ 0.173}}
& 15.230 \footnotesize{$\pm$ 0.026}
& \phantom{1}9.918 \footnotesize{$\pm$ 0.027}
& \phantom{1}9.928 \footnotesize{$\pm$ 0.027} \\
{AUC Top-100} 
& \textbf{15.516 \footnotesize{$\pm$ 0.127}} 
& 14.619 \footnotesize{$\pm$ 0.027}
& \phantom{1}8.416 \footnotesize{$\pm$ 0.024}
& \phantom{1}8.064 \footnotesize{$\pm$ 0.005} \\

\bottomrule
\end{tabular}

}
\end{subtable}
\begin{subtable}{0.27\textwidth}
\centering
\caption{Search distances ($\uparrow$)}\label{tab:dist}
\resizebox{\linewidth}{!}{
\begin{tabular}{lcc}
\toprule
(GSK3$\beta$) & SMILES & Molecule  \\
\midrule
\ours & 0.740 & 0.528 \\
LS-GFN & 0.374 & 0.494 \\
\bottomrule
\toprule
(JNK3) & SMILES & Molecule  \\
\midrule
\ours & 0.706 & 0.536 \\
LS-GFN & 0.403 & 0.512 \\
\bottomrule
\end{tabular} 


}
\end{subtable}
\vspace{-5pt}
\end{table}

\subsubsection{Sample efficient multi-objective molecular optimization} \label{sec:moo}

\begin{wraptable}{r}{0.475\linewidth}
    \centering
    \vspace{-12pt}
    \caption{Average and standard deviation of hypervolumes ($\uparrow$) for each task. The baseline results are directly from the HN-GFN paper \cite{zhu2024sample_moo}. The \textbf{bold} text indicates the best value.}
    \resizebox{\linewidth}{!}{\begin{tabular}{lcc}
\toprule
 & GSK3$\beta$ + JNK3 & \makecell{GSK3$\beta$ + JNK3 \\+ QED + SA}  \\
\midrule
HierVAE+$q$ParEGO & 0.205 \footnotesize{$\pm$ 0.015} & 0.186 \footnotesize{$\pm$ 0.009} \\
HierVAE+$q$EHVI & 0.341 \footnotesize{$\pm$ 0.072} & 0.211 \footnotesize{$\pm$ 0.006} \\
LaMOO & 0.279 \footnotesize{$\pm$ 0.090} &  0.190 \footnotesize{$\pm$ 0.069} \\
Graph GA & 0.368 \footnotesize{$\pm$ 0.020} & 0.335 \footnotesize{$\pm$ 0.021} \\
MARS & 0.418 \footnotesize{$\pm$ 0.095} & 0.273 \footnotesize{$\pm$ 0.020} \\
HN-GFN & 0.669 \footnotesize{$\pm$ 0.061} & 0.416 \footnotesize{$\pm$ 0.023} \\
\midrule
\ours{} & \textbf{0.718 \footnotesize{$\pm$ 0.138}} & \textbf{0.642 \footnotesize{$\pm$ 0.053}} \\
\bottomrule
\end{tabular} }
    \label{tab:moo}
\end{wraptable}
According to Zhu et al. \cite{zhu2024sample_moo}, we apply our method to multi-objective tasks: GSK3$\beta$+JNK3 and GSK3$\beta$+JNK3+QED+SA. Notably, GSK3$\beta$ and JNK3 are potential targets of Alzheimer’s Disease treatments \cite{li2018jnk_gsk3b}. We use a linear combination of each objective with given coefficients, and the performance is measured by hypervolumes with 1K evaluations.
We obtained the results from five independent trials using different seeds.
Even though \ours{} is not designed for multi-objective molecular optimization, it demonstrates notable performance using proper scalar-valued score functions; please see \cref{appnd:moo} for details.

\subsubsection{Further analysis}
\textbf{Active learning with \ours{}.} Similar to GFlowNet-AL, ours can work as a generative model in multi-round active learning. We compare \ours{}-AL with other model-based and active learning methods; please refer to \cref{appnd:al_exp}.

\textbf{\ours{} with SELFIES representation.} \ours{} with SELFIES generation achieves the improved sample efficiency to other SELFIES-based methods; see \cref{appnd:selfies}.

\textbf{Sensitivity analysis.} We provide the experimental results by varying the hyperparameters, such as the offspring size and the number of training loops; please refer to \cref{appnd:sensistivity}.

\subsection{Designing inhibitors against SARS-CoV-2 targets} \label{sec:exp_covid}

In this subsection, we conduct drug discovery experiments for Severe Acute Respiratory Syndrome Coronavirus 2 (SARS-Cov-2), known as the novel coronavirus. One desired property is maximizing the binding affinity to the target protein. The binding affinity is measured with a docking score, which is calculated based on the energies of the interaction between the ligand and the receptor. Typically, the computation of docking scores is expensive since it involves predicting the spatial orientation and binding affinity of the molecule in the active site of the target protein. We employ Quick Vina 2 \cite{alhossary2015qvina} docking software to assess generated molecules.

Additionally, QED (Quantitative Estimate of Drug-likeness) and SA (Synthetic Accessibility) are considered to quantify the drug-likeness and difficulties of synthesizing. The higher QED, which ranges [0, 1], and the lower SA, which ranges [0, 10], are desired. Therefore, we define the score function $s(x)$ for designing SARS-Cov-2 inhibitors as a linear combination of normalized scores according to the previous work \cite{hu2024molrl}. Following \cite{rogers2023sars} and \cite{hu2024molrl}, the target proteins are selected: {PLPro\_7JIR}, a critical enzyme in the life cycle of SARS-CoV-2, and RdRp\_6YYT, which is essential for the replication and the transcription of genes.

\begin{figure}
\centering
\includegraphics[width=0.975\textwidth]{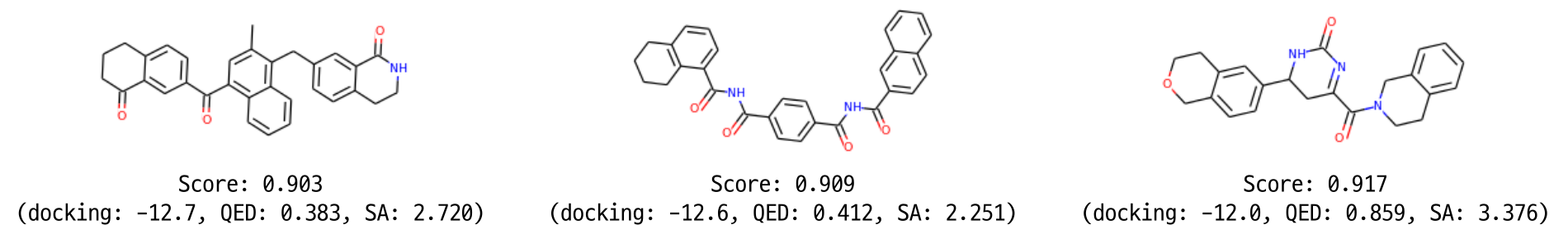}
\caption{The final candidates for the PLPr\_7JIR target with 100 steps. 
} \label{fig:plpro}
\end{figure}

\begin{figure}
\vspace{-5pt}
\centering
\includegraphics[width=0.975\textwidth]{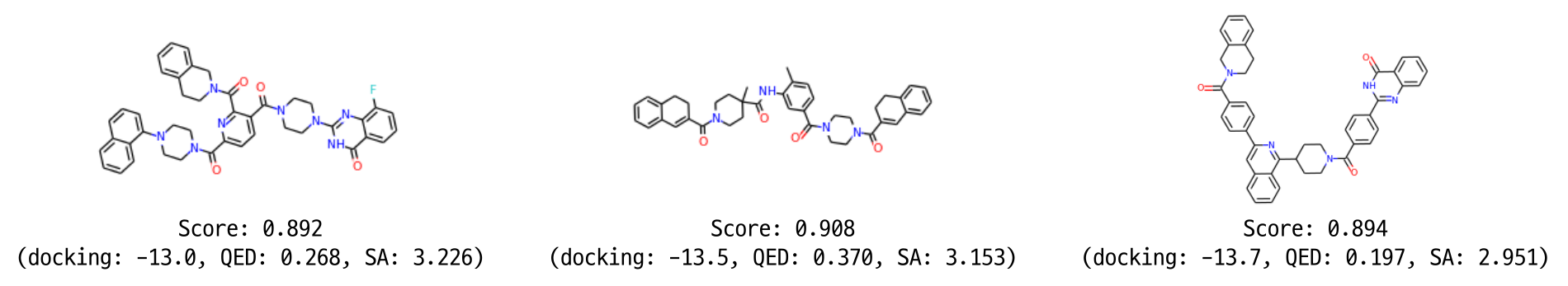}
\caption{The final candidates for the RdRp\_6YYT target with 100 steps. 
} \label{fig:rdrp}
\end{figure}

\begin{wraptable}{r}{0.42\linewidth}
    \centering
    \vspace{-15pt}
    \caption{Average Top-100 scores ($\uparrow$). Ours outperforms baselines with 10 times fewer steps. The \textbf{bold} denotes the best scores.} \label{tab:sas-cov}
    \vspace{-5pt}
    \resizebox{0.85\linewidth}{!}{\begin{tabular}{lcc}
\toprule
 & PLPro & RdRp  \\
\midrule
JT-VAE & 0.272 & 0.216 \\
GFlowNet & 0.326 & 0.280 \\
Graph GA & 0.723 & 0.786 \\
REINVENT & 0.717 & 0.799 \\
MolRL-MGPT & 0.772 & 0.854 \\
\midrule
\ours{} (100) & 0.891 & 0.873 \\
\ours{} (1000) & \textbf{0.925} & \textbf{0.902} \\
\bottomrule
\end{tabular} }
\end{wraptable}
The experiments are conducted with up to 1000 update steps with 128 batch size \cite{hu2024molrl}. As shown in \cref{tab:sas-cov}, ours achieves the highest Top-100 average scores only with 100 steps, which is 10 times fewer than others. Note that the score is recalculated based on the normalized score function in \cref{eq:sas_cov} using average values in the MolRL-MGPT paper \cite{hu2024molrl}; the full results and a more detailed experimental setup are provided in \cref{appnd:sas_cov}.
We also report the best candidates of 100 steps in \cref{fig:plpro} and in \cref{fig:rdrp}. The final molecules correspond to the Top-1 score molecules from 3 independent runs.

\section{Discussion}

This paper introduces a Genetic-guided GFlowNet (\ours{}), which integrates a domain-specific genetic algorithm to guide the GFlowNet policy toward higher-reward samples. The method employs off-policy training with a rank-based reweighted buffer, enhancing the policy as a powerful amortized inference sampler for chemical discovery. Extensive experiments demonstrate that \ours{} effectively generates desirable molecules within the high-dimensional chemical space, including long chemical structure sequences (e.g., $\geq$ 100). On the other hand, our approach can be considered as a novel population reinitialization strategy for genetic algorithms using GFlowNets, which sample diverse objects proportional to rewards.

\textbf{Limitations and future works.}
Our method assumes the existence of effective genetic algorithms, which is valid for the molecular design domain. However, designing domain-specific operators for genetic algorithms can be challenging in other fields. 
One possible future work is to enhance genetic algorithms using the neural policy similar to recent studies \cite{fu2022reinforced_ga,nigam2022janus}.
Another direction is to extend our approach to other domains, such as combinatorial optimization. For instance, we could utilize a powerful GA, hybrid genetic search \cite{vidal2012hybrid}, to design a GFlowNet-based solver for routing problems.

\textbf{Broader Impact.} This paper introduces a new generative model, significantly enhancing sample efficiency in molecular optimization. This advance is likely to hold substantial promise for this field, potentially accelerating the development of new therapies and advanced materials. Our research is currently focused on \textit{in-silico} experiments. The potential safety concerns of discovered molecules are further examined in the subsequent processes, such as \textit{in-vitro} experiments and pre-clinical tests.

\begin{ack}
This work was supported by the National Research Foundation of Korea (NRF) grant funded by the Korea government (MSIT) (No. RS-2024-00410082). The authors are grateful to Austin Tripp and Xiuyuan Hu for
their help with baselines.
\end{ack}


\bibliographystyle{unsrt}
\bibliography{references}

\newpage
\appendix

\clearpage
\section{Implementation details of \ours{}}

\ours{} is implemented on top of the PMO benchmark source code (MIT license).\footnote{\href{https://github.com/wenhao-gao/mol_opt}{https://github.com/wenhao-gao/mol\_opt}} Mostly, we adopt the REINVENT implementation including the RNN models and experience buffer; the code is included in the benchmark, and the original implementation is also accessible with Apache-2.0 license.\footnote{\href{https://github.com/MolecularAI/Reinvent}{https://github.com/MolecularAI/Reinvent}} See the following subsections for details.

\begin{figure}
    \centering
    \includegraphics[width=0.92\textwidth]{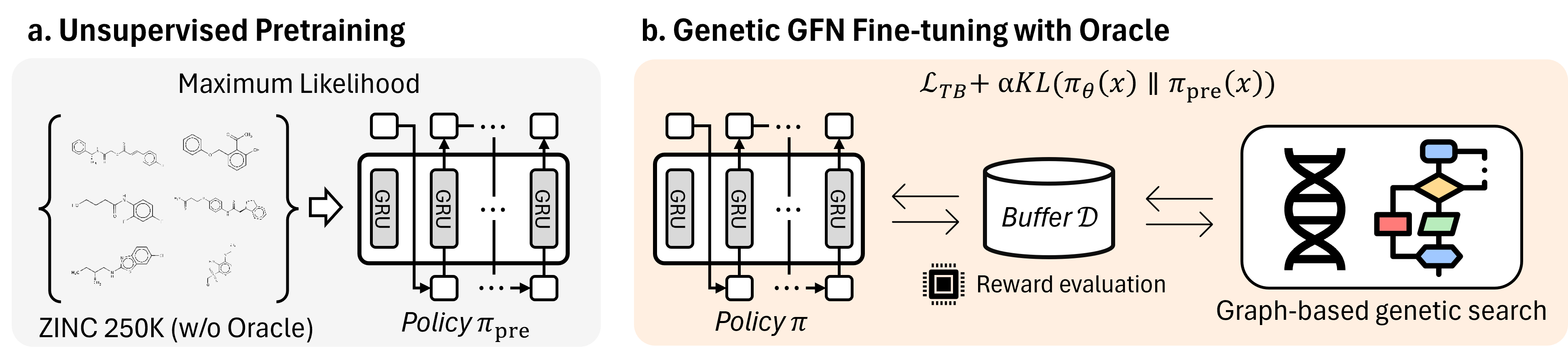}
    \caption{Pretraining and Genetic GFN fine-tuning framework} \label{fig:pretraining}
\end{figure}

\subsection{Network architecture and pretraining} \label{appnd:pre-training}

\paragraph{Network architecture.} Our policy network is parameterized using a recurrent neural network containing multiple GRU cells \citep{chung2014gru}. In molecular optimization, RNN-based models with string molecular representations have proven to be successful \cite{olivecrona2017molecular, ahn2020guiding, hu2024molrl}. In experiments, we employ the same hyperparameters to directly compare with REINVENT, whose input embedding dimension is 128 and hidden dimension is 512 with three layers. 

\paragraph{Pretraining.} According to the PMO benchmark guidelines \citep{gao2022sample}, the pre-training is conducted on ZINC 250K. The overall framework is illustrated in \cref{fig:pretraining}. Since the network architecture is the same as REINVENT, we adopt the provided pretrained model for REINVENT in the PMO benchmark. This allows the direct comparison of fine-tuning approaches.

\subsection{Hyperparameters} 
We mostly follow the hyperparmeter setup of REINVENT and GEGL. For instance, the batch size and learning rate are set as 64 and 0.0005 according to REINVENT in the PMO benchmark. On the other hand, the mutation rate and the number of training loops are set to 0.01 and 8 following GEGL. We use 64 samples for the replay training and population size, the same as the batch size without tuning. Lastly, the learning rate of $Z$, the partition function, is set to 0.1, also without tuning.

In contrast, we have searched several hyperparameters, offspring size, the number of GA generations, and KL-divergence coefficient $\alpha$. We provide the sensitivity analysis for the offspring size and the number of GA generations in \cref{appnd:sensistivity}. Furthermore, we use the inverse temperature $\beta=10$ and the weight-shifting factor $k=0.01$, but they can be differently used to control score-diversity trade-off, as explained in \cref{sec:tradeoff_exp}.

\subsection{Computing resource}
Throughout the experiments, we utilize a 48-core CPU, Intel(R) Xeon(R) Gold 5317 CPU @ 3.00GHz, and a single GPU. In the PMO benchmark, runtime varies from less than 10 minutes to several hours for 10K evaluations, depending on tasks and algorithms. However, most of the runtime is consumed in evaluating score functions—the motivation for why the sample efficiency matters. On the other hand, in the SARS-CoV-2 inhibitor design tasks, 1000 training steps with a batch size of 128 require more than 1 day for PlPro\_7JIR and 2 days for RdRp\_6YYT; more than 95\% of the time is used to evaluation \cite{hu2024molrl}.

\clearpage
\section{Genetic operations} \label{appd:genetic}

\begin{figure}
    \centering
    \begin{subfigure}[b]{0.51\textwidth}
    \includegraphics[width=\textwidth]{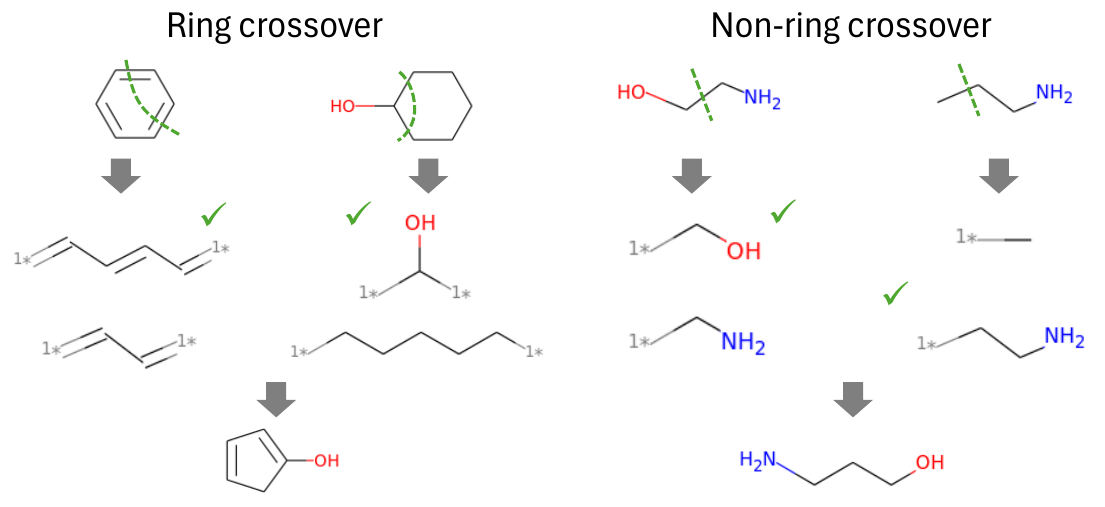}
    \caption{Crossover}
    \end{subfigure}
    \hfill
    \begin{subfigure}[b]{0.46\textwidth}
    \includegraphics[width=\textwidth]{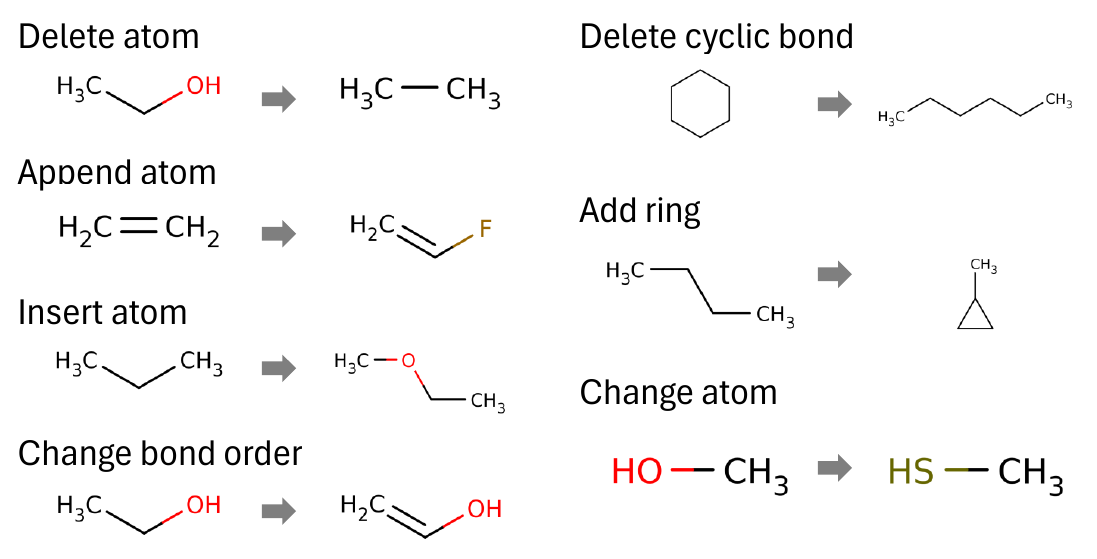}
    \caption{Mutations} 
    \end{subfigure}
    \caption{Examples of Graph GA operations. These operations are conducted according to predefined SMARTS patterns to ensure molecule validity, such as adherence to valence rules.} \label{fig:ga}
\end{figure}

This section details each operation in our genetic search; see \cref{fig:ga} for illustration. Note that we adopt Graph GA of \cite{jensen2019graph}, which has demonstrated its powerful performances and has been adopted by GA-related works like Mol GA \citep{tripp2023genetic} and GEGL \citep{ahn2020guiding}.

\subsection{Crossover}

A crossover operation is conducted to generate a new candidate (called offspring) by exchanging the genetic information of a pair of selected individuals (parents). This process mimics the crossover of genetic material in biological reproduction.
In the context of molecular optimization with graphs, the crossover operation is conducted in two types: `ring crossover' and `non-ring crossover with a 50\% probability.

These two main crossover operations perform crossover between two parent molecules by cutting and recombining ring substructures. Ring crossover performs a ring cut specifically designed to target ring structures within the molecule. 
The ring-cut operation cuts the molecule along two different ring patterns, selected randomly.
One of the ring patterns checks for a specific arrangement of four consecutive ring atoms, and the other pattern checks for a ring atom connected to two other ring atoms with a single bond.
If a suitable ring pattern is found, it cuts the molecule along that pattern, resulting in two fragments. 
On the other hand, non-ring crossover cuts a single bond, meaning it is not part of a cyclic (ring) structure within the molecule.
The obtained fragments from both parents are recombined to create new molecules by applying predefined reaction SMARTS patterns. These operations are repeated for validity to ensure that the resulting molecules meet structural and size constraints.

\subsection{Mutation}

The mutation is a random change that is introduced to the genetic information of some individuals. This step adds diversity to the population and helps explore new regions of the solution space.
In this work, we employ seven different mutation processes and randomly select one of these mutations to modify the offspring molecules slightly.
The operations consist of atom and bond deletions, appending new atoms, inserting atoms between existing ones, changing bond orders, deleting cyclic bonds, adding cyclic rings, and altering atom types.

\begin{enumerate}
    \item Deletion of atom: it selects one of five deletion SMARTS patterns, each representing the removal of a specific number of atoms or bonds. These patterns include the removal of a single atom, a single bond, a bond with two attached atoms, and bonds with multiple attached atoms. The selected pattern is applied to the molecule, deleting the specified atom(s) or bond(s).

    \item Appending atom: it introduces a new atom to the molecule. The type of atom (e.g., C, N, O) and the type of bond (single, double, or triple) are chosen based on predefined probabilities. The function then generates a reaction SMARTS pattern to append the selected atom to the molecule, forming a new bond.

    \item Inserting atom: it inserts a new atom between two existing atoms in the molecule. Similar to the appending atom, it selects the type of atom and bond based on predefined probabilities and generates a reaction SMARTS pattern to insert the atom.

    \item Changing bond order: it randomly selects one of four SMARTS patterns, each representing a change in the bond order between two atoms. These patterns include changing a single bond to a double bond, a double bond to a triple bond, and vice versa.

    \item Deletion of cyclic bond: it deletes a bond that is a part of a cyclic structure within the molecule. The SMARTS pattern represents the breaking of a cyclic bond while retaining the atoms connected by the bond.

    \item Adding ring: it introduces a new cyclic ring into the molecule by selecting one of four SMARTS patterns, each representing the formation of a specific ring type. These patterns create different types of cyclic structures within the molecule.

    \item Changing atom: it randomly selects two types of atoms from a predefined list and generates a SMARTS pattern to change one atom into another. This operation modifies the atom type within the molecule.
\end{enumerate}

\clearpage
\section{Hyperparameter setup for baseline methods} \label{appnd:tuning}
In this subsection, we present detailed descriptions for hyperparameter tuning of baselines. 
Except for Mol GA and GEGL, we adopt all the hyperparameters, initial datasets, and pre-trained models provided in the PMO benchmark.
For hyperparameters that affect the sample efficiency, such as population size, we have searched for the proper hyperparameters following the guidelines suggested by \cite{gao2022sample}. In detail, we tune Mol GA and GEGL on the average AUC Top-10, the main performance metric, from 3 independent runs of two oracles, \texttt{zaleplon\_mpo} and \texttt{perindopril\_mpo}. The best configurations are used in the main experiments. 

\begin{figure}[ht!]
    \centering
    \includegraphics[width=0.85\linewidth]{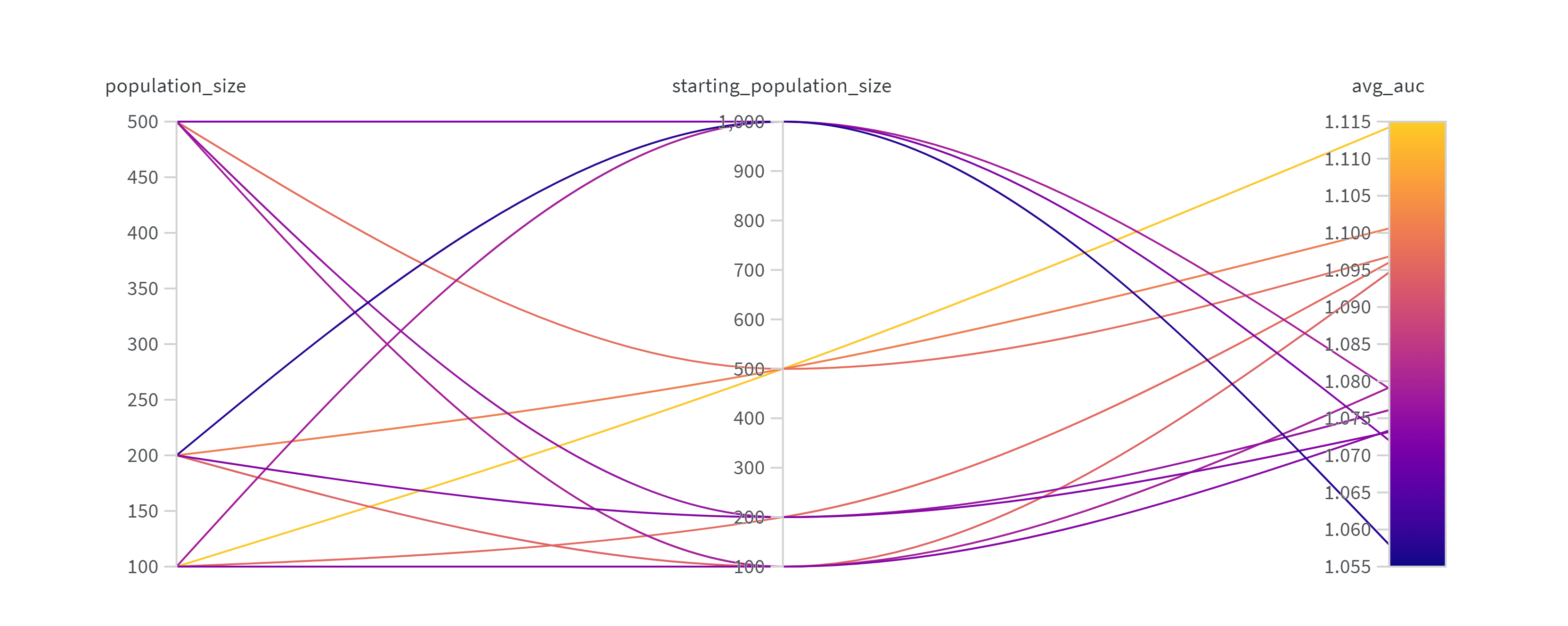}
    \caption{Hyperparameter tuning results for Mol GA}
    \label{fig:mol_ga_sweep}
\end{figure}

\paragraph{Mol GA.} As mentioned in \cref{sec:exp_pmo}, we set the offspring size as 5, the most crucial hyperparameter, according to the original paper \cite{tripp2023genetic}. Then, we searched the starting population size in [100, 200, 500, 1000] and the population size [100, 200, 500]. As shown in \cref{fig:mol_ga_sweep}, we found the best configuration to be 500 and 100 for the starting population size and the population size, respectively.

\begin{figure}[ht!]
    \centering
    \includegraphics[width=0.85\linewidth]{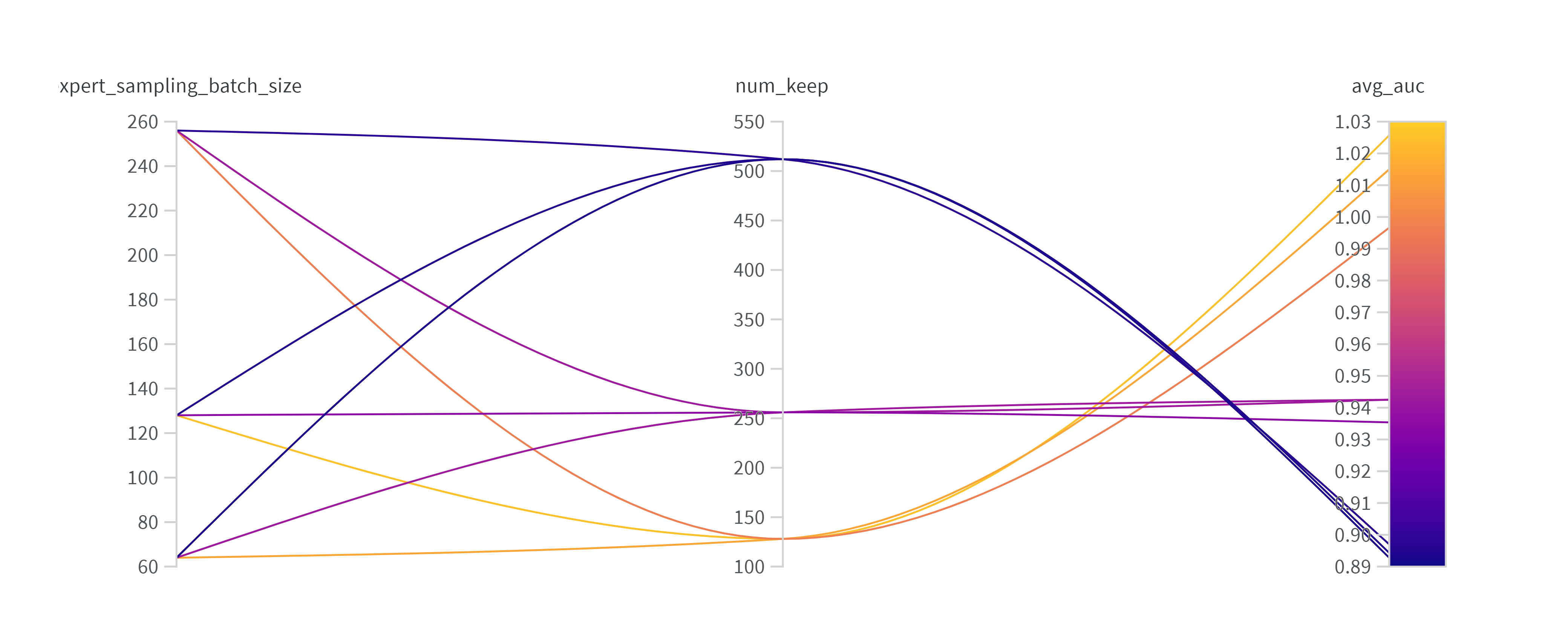}
    \caption{Hyperparameter tuning results for GEGL}
    \label{fig:gegl_sweep}
\end{figure}

\paragraph{GEGL.} In GEGL, the policy sampling size is set as the expert sampling size, and both priority queue sizes (denoted as \texttt{num\_keep}) are the same as the original implementation. Thus, we searched the expert sampling size in [64, 128, 512] and the priority queue size in [128, 512, 1024]. Originally, they were set as 8192 and 1024, which are improper to the sample efficient setting. For the training batch size, we use 64, which is the same as ours.
We use the pretrained policy provided in the original code and adapt the setup of the rest of the hyperparameters, including mutation rate, learning rate, and the number of training loops. Based on the results in \cref{fig:gegl_sweep}, we set the expert sampling size and priority queue size as 128.

\clearpage
\section{Comparison Genetic GFN-AL with active learning and model-based algorithms} \label{appnd:al_exp}

\subsection{Implementation of Genetic GFN-AL} \label{appnd:al}

We mostly adopt the implementation of GFlowNet-AL, including the training method and the utilization of an acquisition function \cite{jain2022biological}. The source code is accessible online with an MIT license.\footnote{\href{https://github.com/MJ10/BioSeq-GFN-AL}{https://github.com/MJ10/BioSeq-GFN-AL}}
Additionally, the number of training proxy and generative models are aligned with the PMO benchmark standards, while the replay training and genetic search hyperparameters are set to those used in our method in \ours{}. The pseudo-code is as follows.

\begin{algorithm}[ht]
   \caption{Multi-round active learning with \ours{}} \label{alg:ours-al}
\begin{algorithmic}[1]
    \Require Pretrained policy $\pi_{\text{pre}}$
    \Ensure Top-$K$ elements of discovered molecule dataset $\mathcal{D}$
    \State Set $\pi_\theta \gets \pi_{\text{pre}}$, $\mathcal{D} \gets \emptyset$ 
    \State Initialize the proxy model $f_\phi$
    \While{$|\mathcal{D}| \leq \texttt{numOracle}$}
    \State \textcolor{purple}{\(\triangleright\) \quad \textsc{Proxy training}}
    \For{$k=1$ {\bfseries to} \texttt{numTrainProxy}}
        \State Sample $\bm{x}$ from $\pi_\theta$ with \texttt{GeneticSearch}
        \State $y^{(i)} \gets \mathcal{O}(\bm{x}^{(i)})$
        \State $\mathcal{D} \gets \mathcal{D} \cup \{(\bm{x}^{(i)}, y^{(i)})\}_{i=0}^n$
        \State Update $\phi$ to minimize $\sum_{(\bm{x}, y) \in \mathcal{D}} (f_\phi(\bm{x}) - y)^2$
    \EndFor
    \State \textcolor{purple}{\(\triangleright\) \quad \textsc{Generative policy training}}
    \State $\mathcal{D}_{\text{inner}} \gets \emptyset$
    \For{$l=1$ {\bfseries to} \texttt{numTrainPolicy}}
        \State Get $m'=\lceil m(1-\gamma) \rceil$ samples from $\pi_\theta$ with \texttt{GeneticSearch}
        \State Get $m - m'$ samples from $\mathcal{D}$
        \State $\mathcal{D}_{\text{inner}} \gets \mathcal{D}_{\text{inner}} \cup \{(\bm{x}^{(i)}, \mathcal{F}(\mu(\bm{x}^{(i)}), \sigma(\bm{x}^{(i)})) \}_{i=0}^m$ \Comment{Acquisition function from \cite{jain2022biological}}
    
        \For{$r=1$ {\bfseries to} \texttt{numReplay}}
        \State Get $\mathcal{B}$ from $\mathcal{D}_{\text{inner}}$ with rank-based sampling
        \State Update $\theta$ to minimize $\frac{1}{|\mathcal{B}|}\sum_{(\bm{x}, f_\phi(\bm{x}))\in \mathcal{B}}\mathcal{L}_{\text{TB}} + \alpha \text{KL}(\pi_\theta(\bm{x})||\pi_{\text{pre}}(\bm{x}))$
        \EndFor
    \EndFor
    \EndWhile
\end{algorithmic}
\end{algorithm}

\paragraph{Hyperparmeters.} We use the same hyperparameters for the generative model (i.e., \ours{}). Since active learning approaches introduce various hyperparameters, such as the training iterations of the proxy and generative models, not only introduce the proxy model, we tried to keep the setup of GFlowNet-AL in the PMO benchmark (e.g., proxy learning rate). We provide hyperparameters in \cref{tab:al_hyper}; note that our proxy model predicts the score using SMILES rather than fragments, unlike the original GFlowNet-AL. Additionally, we adopt $\gamma$, the ratio of offline (oracle-touched) data in the inner loop training, and the acquisition function (related with $\kappa$); see the original GFlowNet-AL paper \cite{jain2022biological}.

\begin{table}
    \centering
    \caption{Hyperparameters of \ours{}-AL} \label{tab:al_hyper}
    \resizebox{0.7\linewidth}{!}{
\begin{tabular}{lcc}
\toprule
     & \makecell{SMILES\\ \ours{}-AL} & \makecell{Fragment\\GFlowNet-AL\\(PMO benchmark)} \\
    \midrule
    proxy init sample size & 480 & 500 \\
    proxy sample size & 480 & 500 \\
    proxy training iterations & 25 & 25 \\
    proxy training batch size & 64 & 64 \\
    proxy hidden dimension & 512 & 64 \\
    proxy layers & 3 & 3 \\
    generative model training iterations & $8 \times 10$ & 100 \\
    proxy learning rate & 0.00025 & 0.00025 \\
    proxy weight decay & 0.000001 & 0.000001 \\
    proxy dropout & 0.0 & 0.0 \\
    kappa & 0.1 & 0.0 \\
    gamma & 0.5 & 0.0 \\
    random action prob & 0.0 & 0.05 \\
\bottomrule
\end{tabular}}
\end{table}

\subsection{Experimental results}

We compare \ours{}-AL with various model-based and active learning methods. Here, GP BO is regarded as a model-based method of Graph GA because GP BO uses Graph when optimizing the GP acquisition function.
As shown in \cref{tab:al}, SMILES \ours{}-AL achieves significantly improved performance compared to fragment GFlowNet-AL in the PMO benchmark. It is noteworthy that we further utilize the acquisition function, not directly use the proxy prediction as a reward, and mix the oracle-touched data, as described in the previous section. Therefore, to verify the effectiveness of genetic search in the active learning setup, we implement another baseline, SMILES GFN-AL using $\epsilon$-greedy exploration, similar to our self-ablation studies in \cref{sec:tradeoff_exp}. The results demonstrate that \ours{} is beneficial to enhancing sample efficiency in the active learning setting. As pointed out in the PMO benchmark, though the model-based methods (or active learning methods) are known as more sample efficient, they require careful design to achieve superior performances.

\begin{table}[ht]
\centering
\caption{Comparing AL and model-based algorithms. The \textbf{bold} text indicates the best value.}
\resizebox{0.85\linewidth}{!}{\begin{tabular}{lcccc}
\toprule
 & \ours{}-AL & \makecell{SMILES GFN-AL \\ with $\eps$-greedy} & GP BO \cite{tripp2021fresh} & \makecell{Fragments\\GFlowNet-AL \cite{jain2022biological}} \\
\midrule
AUC Top-1
& \textbf{13.997} \footnotesize{$\pm$ 0.337} 
& 11.954 \footnotesize{$\pm$ 0.190}
& 13.718 \footnotesize{$\pm$ 0.080}
& 11.032 \footnotesize{$\pm$ 0.016} \\
AUC Top-10 
& \textbf{13.462} \footnotesize{$\pm$ 0.303}
& 11.257 \footnotesize{$\pm$ 0.171}
& 13.115 \footnotesize{$\pm$ 0.074}
& \phantom{1}9.928 \footnotesize{$\pm$ 0.027} \\
AUC Top-100 
& \textbf{12.263} \footnotesize{$\pm$ 0.218}  
& \phantom{1}9.901 \footnotesize{$\pm$ 0.237}
& 12.050 \footnotesize{$\pm$ 0.082}
& \phantom{1}8.064 \footnotesize{$\pm$ 0.005}\\
\bottomrule
\end{tabular}
}
\label{tab:al}
\end{table}

\clearpage
\section{Multi-objective sample efficient molecular optimization tasks} \label{appnd:moo}

This section provides details on multi-objective sample efficient molecular optimization experiments. Even though \ours{} targets single (i.e., scalar-valued) objective optimization tasks, ours directly applies to multi-objective tasks using well-defined coefficients. Inspired by the work of Zhu et al. \cite{zhu2024sample_moo}, we conduct experiments on two multi-objective tasks: GSK3$\beta$ + JNK3 and GSK3$\beta$ + JNK3 + QED + SA. \ours{} is implemented on top of the original code (MIT license).\footnote{\href{https://github.com/violet-sto/HN-GFN}{https://github.com/violet-sto/HN-GFN}} 

Interestingly, GSK3$\beta$ and JNK3 are machine-learning-based oracles that estimate inhibiting scores against the Glycogen synthase kinase 3 beta target and the c-Jun N-terminal kinase 3 target. Designing dual inhibitors for both targets can be beneficial to designing treatments for Alzheimer’s Disease \cite{li2018jnk_gsk3b}. 
We define scalar-valued score functions using the linear combination of oracle functions with given coefficients. The cost coefficients $\alpha$ are set following the HN-GFN paper, i.e., $\alpha=(1, 1)$ for GSK3$\beta$ + JNK3 and $\alpha=(3, 4, 2, 1)$ for GSK3$\beta$ + JNK3 + QED + SA. Since the reward scales become larger, we reduce the inverse temperature from 50 to 25.

According to the experiment setup of hypernetwork-based GFlowNets (HN-GFN) \cite{zhu2024sample_moo}, we limit the reward calls to 1000. Consequently, the batch size, replay training loops, population, and offspring size are adjusted to half. The performance is measured using hypervolumes with five independent runs. Note that we use the same pretrained model in the main experiment for PMO.

\clearpage
\section{Designing of SAS-Cov-2 inhibitors} \label{appnd:sas_cov}

According to the previous work \cite{hu2024molrl}, we define the score function $s(x)$ for designing SARS-Cov-2 inhibitors as a linear combination of normalized scores, i.e., 

\begin{equation} \label{eq:sas_cov}
    s(x) = 0.8 \cdot \frac{1}{1+10^{0.625 (s_{\text{docking}}(x)+10)}} + 0.1 \cdot s_{\text{QED}}(x) + 0.1 \cdot \frac{10 - s_{\text{SA}}(x)}{9}.
\end{equation}

The target proteins are as follows:
\begin{itemize}
    \item \textbf{PLPro\_7JIR}: PLPro (papain-like protease) is a critical enzyme in the life cycle of SARS-CoV-2, which can help in studying the enzyme's function and in designing inhibitors that could potentially disrupt the virus’s ability to replicate and evade the immune system. The 7JIR represents a C111S mutant version of PLPro.\footnote{\href{https://www.rcsb.org/structure/7JIR}{https://www.rcsb.org/structure/7JIR}} %
    \item \textbf{RdRp\_6YYT}: RdRp (RNA-dependent RNA polymerase) is essential for the replication of the genome and the transcription of genes in SARS-CoV-2. The protein structure of RdRp is cataloged in the Protein Data Bank (PDB) under the identification code 6YYT.\footnote{\href{https://www.rcsb.org/structure/6YYT}{https://www.rcsb.org/structure/6YYT}}
\end{itemize}

\ours{} is implemented on top of the implementation of MolRL-MGPT (Molecular design using Reinforcement Learning with Multiple GPT agents).\footnote{Available at \href{https://github.com/HXYfighter/MolRL-MGPT}{https://github.com/HXYfighter/MolRL-MGPT}} We employ the same hyperparameters with the experiments on the PMO benchmark, except for the weight-shifting factor (we use $k=0.05$).

In \cref{tab:sas-cov}, the scores of baselines are computed according to the \cref{eq:sas_cov} using the average values in the MolRL-MGPT \cite{hu2024molrl}. We also provide average scores and standard deviation of docking, QED, SA, and diversity in \cref{tab:plpro} and \cref{tab:rdrp}.

\begin{table}[ht]
    \centering
    \caption{The results of Top-100 molecules for PLPr\_7JIR target. The docking, QED, and SA of baselines are directly from MolRL-MGPT, and the total score is recalculated according to \cref{eq:sas_cov} using average values.} \label{tab:plpro}
    \resizebox{0.9\textwidth}{!}{\begin{tabular}{lccccc}
\toprule
 & Score ($\uparrow$) & Docking ($\downarrow$) & QED ($\uparrow$) & SA ($\downarrow$) & Diversity ($\uparrow$) \\
 \midrule
JT-VAE & 0.272 & -8.76 \footnotesize{$\pm$ 0.35} & 0.795 \footnotesize{$\pm$ 0.038} & 2.994 \footnotesize{$\pm$ 0.140} & 0.836 \footnotesize{$\pm$ 0.032} \\
GFlowNet & 0.326 & -9.11 \footnotesize{$\pm$ 0.21} & 0.726 \footnotesize{$\pm$ 0.015} & 2.823 \footnotesize{$\pm$ 0.076} & 0.825 \footnotesize{$\pm$ 0.010} \\
Graph GA & 0.723 & -10.83 \footnotesize{$\pm$ 0.08} & 0.380 \footnotesize{$\pm$ 0.013} & 3.638 \footnotesize{$\pm$ 0.162} & 0.740 \footnotesize{$\pm$ 0.017} \\
Reinvent & 0.717 &  -10.75 \footnotesize{$\pm$ 0.05} & 0.392 \footnotesize{$\pm$ 0.008} & 2.649 \footnotesize{$\pm$ 0.035} & 0.619 \footnotesize{$\pm$ 0.023} \\
MolRL-MGPT & 0.772 & -11.02 \footnotesize{$\pm$ 0.06} & 0.386 \footnotesize{$\pm$ 0.006} & 2.550 \footnotesize{$\pm$ 0.047} & 0.745 \footnotesize{$\pm$ 0.008} \\
\midrule
\ours{} (Ours) & 0.908 & -12.86 \footnotesize{$\pm$ 0.17} & 0.425 \footnotesize{$\pm$ 0.092} & 2.819 \footnotesize{$\pm$ 0.105} & 0.592 \footnotesize{$\pm$ 0.010} \\
\bottomrule
\end{tabular}}
\end{table}

\begin{table}[ht]
    \centering
    \caption{The results of Top-100 molecules for RdRp\_6YYT target. The docking, QED, and SA of baselines are directly from MolRL-MGPT, and the total score is recalculated according to \cref{eq:sas_cov} using average values.} \label{tab:rdrp}
    \resizebox{0.9\textwidth}{!}{\begin{tabular}{lccccc}
\toprule
 & Score ($\uparrow$) & Docking ($\downarrow$) & QED ($\uparrow$) & SA ($\downarrow$) & Diversity ($\uparrow$) \\
 \midrule
JT-VAE & 0.216 & -8.33 \footnotesize{$\pm$ 0.25} & 0.719 \footnotesize{$\pm$ 0.019} & 2.959 \footnotesize{$\pm$ 0.094} & 0.828 \footnotesize{$\pm$ 0.018} \\
GFlowNet & 0.280 & -8.89 \footnotesize{$\pm$ 0.16} & 0.656 \footnotesize{$\pm$ 0.033} & 2.854 \footnotesize{$\pm$ 0.061} & 0.770 \footnotesize{$\pm$ 0.015} \\
GraphGA & 0.786 & -11.26 \footnotesize{$\pm$ 0.12} & 0.262 \footnotesize{$\pm$ 0.010} & 3.520 \footnotesize{$\pm$ 0.049} & 0.658 \footnotesize{$\pm$ 0.009} \\
Reinvent & 0.799 & -11.30 \footnotesize{$\pm$ 0.04} & 0.275 \footnotesize{$\pm$ 0.006} & 2.917 \footnotesize{$\pm$ 0.035} & 0.616 \footnotesize{$\pm$ 0.021} \\
MolRL-MGPT & 0.854 & -11.84 \footnotesize{$\pm$ 0.07} & 0.278 \footnotesize{$\pm$ 0.005} & 2.894 \footnotesize{$\pm$ 0.072} & 0.670 \footnotesize{$\pm$ 0.013} \\
\midrule
\ours{} (Ours) & 0.890 & -13.26 \footnotesize{$\pm$ 0.13} & 0.277 \footnotesize{$\pm$ 0.076} & 3.624 \footnotesize{$\pm$ 0.060} & 0.708 \footnotesize{$\pm$ 0.010} \\
\bottomrule
\end{tabular}
}
\end{table}

\begin{figure}[ht!]
    \centering
    \begin{subfigure}[b]{0.475\textwidth}
    \includegraphics[width=\textwidth]{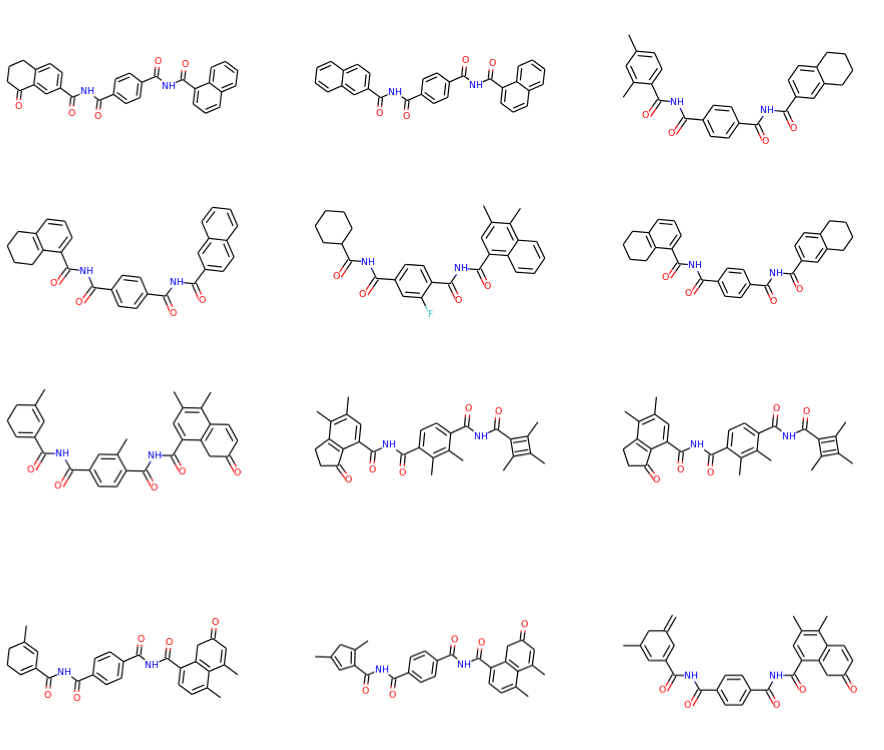}
    \caption{PLPro\_7JIR} 
    \end{subfigure}
    \hfill
    \begin{subfigure}[b]{0.47\textwidth}
    \includegraphics[width=\textwidth]{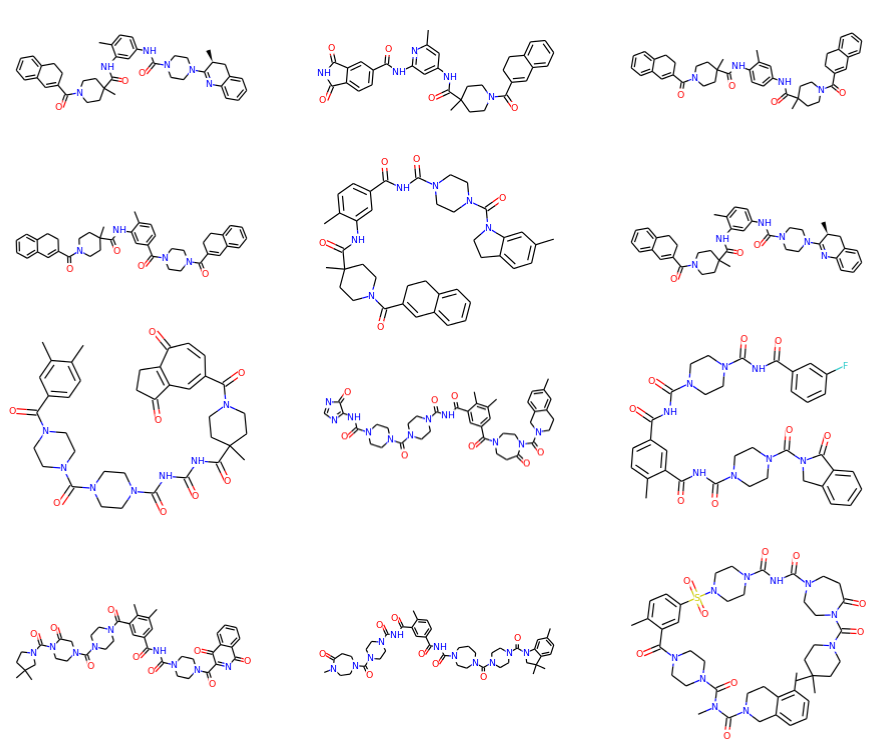}
    \caption{RdRp\_6YYT} 
    \end{subfigure}
    \vspace{-3pt}
    \caption{Examples of Top3 inhibitors for SARS-CoV-2 over steps (\texttt{seed 1})} \label{fig:sars-cov-additonal}
\end{figure}

We provide additional visual results in \cref{fig:sars-cov-additonal}. There seems to be a trend of increasing molecular complexity and functional diversity over iterations.

\vspace{-5pt}

\clearpage
\section{Additional results}

\subsection{Controllability of the score-diversity using the weight-shifting factor $k$} \label{appnd:k}

\begin{figure}[ht]
    \centering
    \includegraphics[width=.6\linewidth]{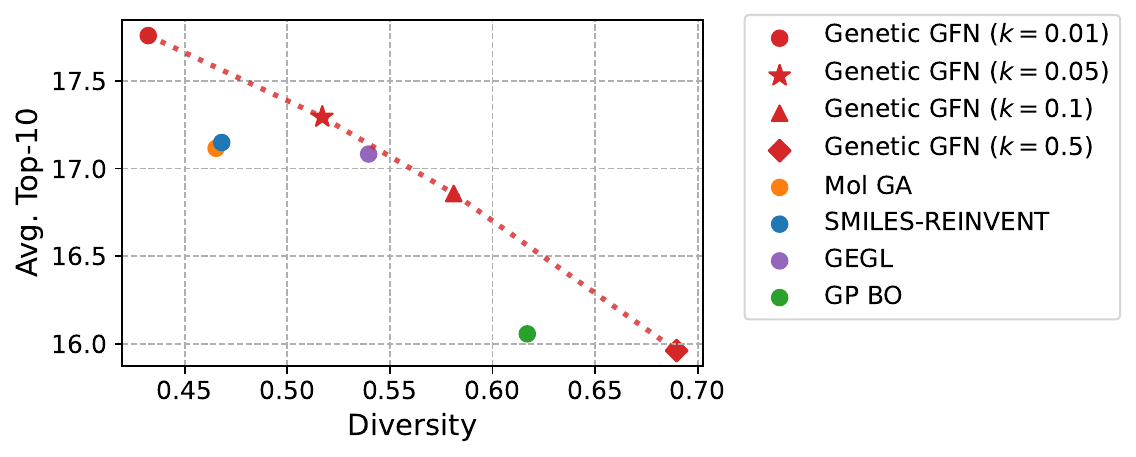}
    \caption{The average score and diversity with adjustments of $k$.}
    \label{fig:k_ablation}
\end{figure}

As mentioned in \cref{sec:tradeoff_exp}, the weight-shifting factor $k$ in \cref{eq:rank} also can control the score-diversity trade-off. Increasing $k$ gives more diverse candidates by increasing the probability of high-ranked samples. Similar to adjusting the inverse temperature, the results in \cref{tab:k_ablation} and \cref{fig:k_ablation} demonstrate that adjusting $k$ with fixed $\beta$ also can effectively control the score-diversity trade-off.

\begin{table}[ht]
    \centering
    \caption{The score-diversity trade-off by varying $k$ with fixed $\beta$.}
    \resizebox{0.6\linewidth}{!}{\begin{tabular}{l|cccc}
\toprule
Oracle &  $k=0.1$  & $k=0.05$ & $k=0.01$ & $k=0.005$   \\

\midrule
\textbf{AUC Top-1} & 15.246 & 15.801 & \textbf{16.527} & 16.323 \\
\textbf{AUC Top-10} & 14.652 & 15.330 & \textbf{16.213} & 16.040 \\
\textbf{AUC Top-100} & 13.597 & 14.453 & \textbf{15.516} & 15.418 \\
\textbf{Diversity} & \textbf{0.581} & 0.515 & 0.432 & 0.444 \\
\bottomrule
\end{tabular} 
}
    \label{tab:k_ablation}
\end{table}

\subsection{Additional results for the PMO benchmark}  \label{appnd:full_pmo}

\subsubsection{Statistical analysis}

We provided the results of the t-tests of AUC Top-10 with Mol GA (2nd place) and SMIELS REINVENT (3rd place). As shown in \cref{tab:t-test_main}, both p-values are less than 0.05, so \ours{} outperforms baselines with statistical significance.

\begin{table}[ht]
    \centering
    \caption{The results of t-tests with Mol GA and REINVENT.}
    \resizebox{0.7\linewidth}{!}{\begin{tabular}{lrrrr}
\toprule
 & \multicolumn{2}{c}{2nd place} & \multicolumn{2}{c}{3rd place} \\
\cmidrule(r){2-3} \cmidrule(r){4-5}
 & \ours & Mol GA & \ours & \makecell{SMILES\\REINVENT} \\
\midrule
Mean & 16.213 & 15.685 & 16.213 & 15.185 \\
Variance & 0.038 & 0.033 & 0.038 & 0.194 \\
Observations & 5 & 5 & 5 & 5 \\
Hypothesized Mean Diff. & 0 & & 0 & \\
t Stat & 4.426 & & 4.780 & \\
P($T \leq t$) one-tail & \textbf{0.001} & & \textbf{0.002} \\
t Critical one-tail & 1.860 & & 2.015 & \\
P($T\leq t$) two-tail & \textbf{0.002} & & \textbf{0.005} \\
t Critical two-tail & 2.306 & & 2.571 \\
\bottomrule
\end{tabular} 
}
    \label{tab:t-test_main}
\end{table}

\clearpage
\subsubsection{Full results of the PMO benchmark}

\begin{table}[ht!]
    \centering
    \caption{Full results of \cref{tab:auc10}.}
    \resizebox{\linewidth}{!}{\begin{tabular}{lccccccc}
\toprule
& \makecell{\ours{}\\(Ours)} & Mol GA & \makecell{SMILES\\ {REINVENT}} & GEGL & GP BO  & \makecell{SELFIES\\ {REINVENT}} & Graph GA \\
\midrule
\#1 & \textbf{0.949 \footnotesize{$\pm$ 0.010}} & 0.928 \footnotesize{$\pm$ 0.015} & 0.881 \footnotesize{$\pm$ 0.016} & 0.842 \footnotesize{$\pm$ 0.019} & 0.902 \footnotesize{$\pm$ 0.011} & 0.867 \footnotesize{$\pm$ 0.025} & 0.859 \footnotesize{$\pm$ 0.013} \\
\#2 & \textbf{0.761 \footnotesize{$\pm$ 0.019}} & 0.740 \footnotesize{$\pm$ 0.055} & 0.644 \footnotesize{$\pm$ 0.019} & 0.626 \footnotesize{$\pm$ 0.018} & 0.579 \footnotesize{$\pm$ 0.035} & 0.621 \footnotesize{$\pm$ 0.015} & 0.657 \footnotesize{$\pm$ 0.022} \\
\#3 & \textbf{0.802 \footnotesize{$\pm$ 0.029}} & 0.629 \footnotesize{$\pm$ 0.062} & 0.717 \footnotesize{$\pm$ 0.027} & 0.699 \footnotesize{$\pm$ 0.041} & 0.746 \footnotesize{$\pm$ 0.025} & 0.588 \footnotesize{$\pm$ 0.062} & 0.593 \footnotesize{$\pm$ 0.092} \\
\#4 & 0.733 \footnotesize{$\pm$ 0.109} & 0.656 \footnotesize{$\pm$ 0.013} & 0.662 \footnotesize{$\pm$ 0.044} & 0.656 \footnotesize{$\pm$ 0.039} & 0.615 \footnotesize{$\pm$ 0.009} & 0.638 \footnotesize{$\pm$ 0.016} & 0.602 \footnotesize{$\pm$ 0.012} \\
\#5 & \textbf{0.974 \footnotesize{$\pm$ 0.006}} & 0.950 \footnotesize{$\pm$ 0.004} & 0.957 \footnotesize{$\pm$ 0.007} & 0.898 \footnotesize{$\pm$ 0.015} & 0.941 \footnotesize{$\pm$ 0.017} & 0.953 \footnotesize{$\pm$ 0.009} & 0.973 \footnotesize{$\pm$ 0.001} \\
\#6 & \textbf{0.856 \footnotesize{$\pm$ 0.039}} & 0.835 \footnotesize{$\pm$ 0.012} & 0.781 \footnotesize{$\pm$ 0.013} & 0.769 \footnotesize{$\pm$ 0.009} & 0.726 \footnotesize{$\pm$ 0.004} & 0.740 \footnotesize{$\pm$ 0.012} & 0.762 \footnotesize{$\pm$ 0.014} \\
\#7 & 0.881 \footnotesize{$\pm$ 0.042} & \textbf{0.894 \footnotesize{$\pm$ 0.025}} & 0.885 \footnotesize{$\pm$ 0.031} & 0.816 \footnotesize{$\pm$ 0.027} & 0.861 \footnotesize{$\pm$ 0.027} & 0.821 \footnotesize{$\pm$ 0.041} & 0.817 \footnotesize{$\pm$ 0.057} \\
\#8 & \textbf{0.969 \footnotesize{$\pm$ 0.003}} & 0.926 \footnotesize{$\pm$ 0.014} & 0.942 \footnotesize{$\pm$ 0.012} & 0.930 \footnotesize{$\pm$ 0.011} & 0.883 \footnotesize{$\pm$ 0.040} & 0.873 \footnotesize{$\pm$ 0.041} & 0.949 \footnotesize{$\pm$ 0.020} \\
\#9 & \textbf{0.897 \footnotesize{$\pm$ 0.007}} & 0.894 \footnotesize{$\pm$ 0.005} & 0.838 \footnotesize{$\pm$ 0.030} & 0.808 \footnotesize{$\pm$ 0.007} & 0.805 \footnotesize{$\pm$ 0.007} & 0.844 \footnotesize{$\pm$ 0.016} & 0.839 \footnotesize{$\pm$ 0.042} \\
\#10 & 0.764 \footnotesize{$\pm$ 0.069} & \textbf{0.835 \footnotesize{$\pm$ 0.040}} & 0.782 \footnotesize{$\pm$ 0.029} & 0.580 \footnotesize{$\pm$ 0.086} & 0.611 \footnotesize{$\pm$ 0.080} & 0.624 \footnotesize{$\pm$ 0.048} & 0.652 \footnotesize{$\pm$ 0.106} \\
\#11 & \textbf{0.379 \footnotesize{$\pm$ 0.010}} & 0.329 \footnotesize{$\pm$ 0.006} & 0.363 \footnotesize{$\pm$ 0.011} & 0.338 \footnotesize{$\pm$ 0.016} & 0.298 \footnotesize{$\pm$ 0.016} & 0.353 \footnotesize{$\pm$ 0.006} & 0.285 \footnotesize{$\pm$ 0.012} \\
\#12 & 0.294 \footnotesize{$\pm$ 0.007} & 0.284 \footnotesize{$\pm$ 0.035} & 0.281 \footnotesize{$\pm$ 0.002} & 0.274 \footnotesize{$\pm$ 0.007} & 0.\textbf{296 \footnotesize{$\pm$ 0.011}} & 0.252 \footnotesize{$\pm$ 0.010} & 0.255 \footnotesize{$\pm$ 0.019} \\
\#13 & 0.708 \footnotesize{$\pm$ 0.057} & \textbf{0.762 \footnotesize{$\pm$ 0.048}} & 0.634 \footnotesize{$\pm$ 0.042} & 0.599 \footnotesize{$\pm$ 0.035} & 0.631 \footnotesize{$\pm$ 0.093} & 0.589 \footnotesize{$\pm$ 0.040} & 0.571 \footnotesize{$\pm$ 0.036} \\
\#14 & \textbf{0.860 \footnotesize{$\pm$ 0.008}} & 0.853 \footnotesize{$\pm$ 0.005} & 0.834 \footnotesize{$\pm$ 0.010} & 0.832 \footnotesize{$\pm$ 0.005} & 0.788 \footnotesize{$\pm$ 0.005} & 0.819 \footnotesize{$\pm$ 0.005} & 0.813 \footnotesize{$\pm$ 0.006} \\
\#15 & 0.595 \footnotesize{$\pm$ 0.014} & \textbf{0.610 \footnotesize{$\pm$ 0.038}} & 0.535 \footnotesize{$\pm$ 0.015} & 0.537 \footnotesize{$\pm$ 0.015} & 0.494 \footnotesize{$\pm$ 0.006} & 0.533 \footnotesize{$\pm$ 0.024} & 0.514 \footnotesize{$\pm$ 0.025} \\
\#16 & \textbf{0.942 \footnotesize{$\pm$ 0.000}} & 0.941 \footnotesize{$\pm$ 0.001} & 0.941 \footnotesize{$\pm$ 0.000} & 0.941 \footnotesize{$\pm$ 0.001} & 0.937 \footnotesize{$\pm$ 0.002} & 0.940 \footnotesize{$\pm$ 0.000} & 0.937 \footnotesize{$\pm$ 0.001} \\
\#17 & 0.819 \footnotesize{$\pm$ 0.018} & \textbf{0.830 \footnotesize{$\pm$ 0.010}} & 0.770 \footnotesize{$\pm$ 0.005} & 0.730 \footnotesize{$\pm$ 0.011} & 0.741 \footnotesize{$\pm$ 0.010} & 0.736 \footnotesize{$\pm$ 0.008} & 0.718 \footnotesize{$\pm$ 0.017} \\
\#18 & \textbf{0.615 \footnotesize{$\pm$ 0.100}} & 0.568 \footnotesize{$\pm$ 0.017} & 0.551 \footnotesize{$\pm$ 0.024} & 0.531 \footnotesize{$\pm$ 0.010} & 0.535 \footnotesize{$\pm$ 0.007} & 0.521 \footnotesize{$\pm$ 0.014} & 0.513 \footnotesize{$\pm$ 0.026} \\
\#19 & 0.634 \footnotesize{$\pm$ 0.039} & \textbf{0.677 \footnotesize{$\pm$ 0.055}} & 0.470 \footnotesize{$\pm$ 0.041} & 0.402 \footnotesize{$\pm$ 0.024} & 0.461 \footnotesize{$\pm$ 0.057} & 0.492 \footnotesize{$\pm$ 0.055} & 0.498 \footnotesize{$\pm$ 0.048} \\
\#20 & \textbf{0.583 \footnotesize{$\pm$ 0.034}} & 0.544 \footnotesize{$\pm$ 0.067} & 0.544 \footnotesize{$\pm$ 0.026} & 0.515 \footnotesize{$\pm$ 0.028} & 0.544 \footnotesize{$\pm$ 0.038} & 0.497 \footnotesize{$\pm$ 0.043} & 0.483 \footnotesize{$\pm$ 0.034} \\
\#21 & \textbf{0.511 \footnotesize{$\pm$ 0.054}} & 0.487 \footnotesize{$\pm$ 0.024} & 0.458 \footnotesize{$\pm$ 0.018} & 0.420 \footnotesize{$\pm$ 0.031} & 0.404 \footnotesize{$\pm$ 0.025} & 0.342 \footnotesize{$\pm$ 0.022} & 0.373 \footnotesize{$\pm$ 0.013} \\
\#22 & 0.135 \footnotesize{$\pm$ 0.271} & 0.000 \footnotesize{$\pm$ 0.000} & \textbf{0.182 \footnotesize{$\pm$ 0.363}} & 0.119 \footnotesize{$\pm$ 0.238} & 0.000 \footnotesize{$\pm$ 0.000} & 0.000 \footnotesize{$\pm$ 0.000} & 0.000 \footnotesize{$\pm$ 0.000} \\
\#23 & \textbf{0.552 \footnotesize{$\pm$ 0.033}} & 0.514 \footnotesize{$\pm$ 0.033} & 0.533 \footnotesize{$\pm$ 0.009} & 0.492 \footnotesize{$\pm$ 0.021} & 0.466 \footnotesize{$\pm$ 0.025} & 0.509 \footnotesize{$\pm$ 0.009} & 0.468 \footnotesize{$\pm$ 0.025} \\
\midrule
Sum & \textbf{16.213}  &  15.686  &  15.185  &  14.354  &  14.264  &  14.152  &  14.131 \\
\midrule
Div. & 0.432 & 0.465 & 0.468 & 0.540 & 0.617 & 0.555 & 0.661 \\
\bottomrule
\end{tabular}
}
    \label{tab:pmo_full}
\end{table}

\begin{table}[ht!]
    \centering
    \caption{Full results of \cref{tab:auc10} (continued).}
    \resizebox{0.95\linewidth}{!}{\begin{tabular}{lcccccc}
\toprule
& \makecell{SMILES\\LSTM-HC} & STONED & \makecell{SynNet} & SMEILS GA & \makecell{Fragment\\GFN} & \makecell{Fragment\\GFN-AL}  \\
\midrule
\#1 & 0.731 \footnotesize{$\pm$ 0.008} & 0.765 \footnotesize{$\pm$ 0.048} & 0.568 \footnotesize{$\pm$ 0.033} & 0.649 \footnotesize{$\pm$ 0.079} & 0.382 \footnotesize{$\pm$ 0.010} & 0.459 \footnotesize{$\pm$ 0.028}  \\
\#2 & 0.598 \footnotesize{$\pm$ 0.021} & 0.608 \footnotesize{$\pm$ 0.020} & 0.566 \footnotesize{$\pm$ 0.006} & 0.520 \footnotesize{$\pm$ 0.017} & 0.428 \footnotesize{$\pm$ 0.002} & 0.437 \footnotesize{$\pm$ 0.007}  \\
\#3 & 0.552 \footnotesize{$\pm$ 0.014} & 0.378 \footnotesize{$\pm$ 0.043} & 0.439 \footnotesize{$\pm$ 0.035} & 0.361 \footnotesize{$\pm$ 0.038} & 0.263 \footnotesize{$\pm$ 0.009} & 0.326 \footnotesize{$\pm$ 0.008}  \\
\#4 & \textbf{0.837 \footnotesize{$\pm$ 0.018}} & 0.612 \footnotesize{$\pm$ 0.007} & 0.635 \footnotesize{$\pm$ 0.043} & 0.612 \footnotesize{$\pm$ 0.005} & 0.582 \footnotesize{$\pm$ 0.001} & 0.587 \footnotesize{$\pm$ 0.002}  \\
\#5 & 0.941 \footnotesize{$\pm$ 0.005} & 0.935 \footnotesize{$\pm$ 0.014} & 0.970 \footnotesize{$\pm$ 0.006} & 0.958 \footnotesize{$\pm$ 0.015} & 0.480 \footnotesize{$\pm$ 0.075} & 0.601 \footnotesize{$\pm$ 0.055}  \\
\#6 & 0.733 \footnotesize{$\pm$ 0.002} & 0.791 \footnotesize{$\pm$ 0.014} & 0.750 \footnotesize{$\pm$ 0.016} & 0.705 \footnotesize{$\pm$ 0.025} & 0.689 \footnotesize{$\pm$ 0.003} & 0.700 \footnotesize{$\pm$ 0.005}  \\
\#7 & 0.846 \footnotesize{$\pm$ 0.019} & 0.666 \footnotesize{$\pm$ 0.022} & 0.713 \footnotesize{$\pm$ 0.057} & 0.714 \footnotesize{$\pm$ 0.038} & 0.589 \footnotesize{$\pm$ 0.009} & 0.666 \footnotesize{$\pm$ 0.006}  \\
\#8 & 0.830 \footnotesize{$\pm$ 0.019} & 0.930 \footnotesize{$\pm$ 0.012} & 0.862 \footnotesize{$\pm$ 0.004} & 0.821 \footnotesize{$\pm$ 0.070} & 0.791 \footnotesize{$\pm$ 0.024} & 0.468 \footnotesize{$\pm$ 0.211}  \\
\#9 & 0.693 \footnotesize{$\pm$ 0.018} & \textbf{0.897 \footnotesize{$\pm$ 0.032}} & 0.657 \footnotesize{$\pm$ 0.030} & 0.853 \footnotesize{$\pm$ 0.049} & 0.576 \footnotesize{$\pm$ 0.021} & 0.199 \footnotesize{$\pm$ 0.199}  \\
\#10 & 0.670 \footnotesize{$\pm$ 0.014} & 0.509 \footnotesize{$\pm$ 0.065} & 0.574 \footnotesize{$\pm$ 0.103} & 0.353 \footnotesize{$\pm$ 0.061} & 0.359 \footnotesize{$\pm$ 0.009} & 0.442 \footnotesize{$\pm$ 0.017}  \\
\#11 & 0.263 \footnotesize{$\pm$ 0.007} & 0.264 \footnotesize{$\pm$ 0.032} & 0.236 \footnotesize{$\pm$ 0.015} & 0.187 \footnotesize{$\pm$ 0.029} & 0.192 \footnotesize{$\pm$ 0.003} & 0.207 \footnotesize{$\pm$ 0.003}  \\
\#12 & 0.249 \footnotesize{$\pm$ 0.003} & 0.254 \footnotesize{$\pm$ 0.024} & 0.241 \footnotesize{$\pm$ 0.007} & 0.178 \footnotesize{$\pm$ 0.009} & 0.174 \footnotesize{$\pm$ 0.002} & 0.181 \footnotesize{$\pm$ 0.002}  \\
\#13 & 0.553 \footnotesize{$\pm$ 0.040} & 0.620 \footnotesize{$\pm$ 0.098} & 0.402 \footnotesize{$\pm$ 0.017} & 0.419 \footnotesize{$\pm$ 0.028} & 0.291 \footnotesize{$\pm$ 0.005} & 0.332 \footnotesize{$\pm$ 0.012}  \\
\#14 & 0.801 \footnotesize{$\pm$ 0.002} & 0.829 \footnotesize{$\pm$ 0.012} & 0.793 \footnotesize{$\pm$ 0.008} & 0.820 \footnotesize{$\pm$ 0.021} & 0.787 \footnotesize{$\pm$ 0.002} & 0.785 \footnotesize{$\pm$ 0.003}  \\
\#15 & 0.491 \footnotesize{$\pm$ 0.006} & 0.484 \footnotesize{$\pm$ 0.016} & 0.541 \footnotesize{$\pm$ 0.021} & 0.442 \footnotesize{$\pm$ 0.016} & 0.423 \footnotesize{$\pm$ 0.006} & 0.434 \footnotesize{$\pm$ 0.006}  \\
\#16 & 0.939 \footnotesize{$\pm$ 0.001} & \textbf{0.942 \footnotesize{$\pm$ 0.000}} & 0.941 \footnotesize{$\pm$ 0.001} & 0.941 \footnotesize{$\pm$ 0.002} & 0.904 \footnotesize{$\pm$ 0.002} & 0.917 \footnotesize{$\pm$ 0.002}  \\
\#17 & 0.728 \footnotesize{$\pm$ 0.005} & 0.764 \footnotesize{$\pm$ 0.023} & 0.749 \footnotesize{$\pm$ 0.009} & 0.723 \footnotesize{$\pm$ 0.023} & 0.626 \footnotesize{$\pm$ 0.005} & 0.660 \footnotesize{$\pm$ 0.004}  \\
\#18 & 0.529 \footnotesize{$\pm$ 0.004} & 0.515 \footnotesize{$\pm$ 0.025} & 0.506 \footnotesize{$\pm$ 0.012} & 0.507 \footnotesize{$\pm$ 0.008} & 0.461 \footnotesize{$\pm$ 0.002} & 0.464 \footnotesize{$\pm$ 0.003}  \\
\#19 & 0.309 \footnotesize{$\pm$ 0.015} & 0.600 \footnotesize{$\pm$ 0.103} & 0.297 \footnotesize{$\pm$ 0.033} & 0.449 \footnotesize{$\pm$ 0.068} & 0.180 \footnotesize{$\pm$ 0.012} & 0.217 \footnotesize{$\pm$ 0.022}  \\
\#20 & 0.440 \footnotesize{$\pm$ 0.014} & 0.375 \footnotesize{$\pm$ 0.029} & 0.397 \footnotesize{$\pm$ 0.012} & 0.310 \footnotesize{$\pm$ 0.019} & 0.261 \footnotesize{$\pm$ 0.004} & 0.292 \footnotesize{$\pm$ 0.009}  \\
\#21 & 0.369 \footnotesize{$\pm$ 0.015} & 0.309 \footnotesize{$\pm$ 0.030} & 0.280 \footnotesize{$\pm$ 0.006} & 0.262 \footnotesize{$\pm$ 0.019} & 0.183 \footnotesize{$\pm$ 0.001} & 0.190 \footnotesize{$\pm$ 0.002}  \\
\#22 & 0.011 \footnotesize{$\pm$ 0.021} & 0.000 \footnotesize{$\pm$ 0.000} & 0.000 \footnotesize{$\pm$ 0.000} & 0.000 \footnotesize{$\pm$ 0.000} & 0.000 \footnotesize{$\pm$ 0.000} & 0.000 \footnotesize{$\pm$ 0.000}  \\
\#23 & 0.470 \footnotesize{$\pm$ 0.004} & 0.484 \footnotesize{$\pm$ 0.015} & 0.493 \footnotesize{$\pm$ 0.014} & 0.470 \footnotesize{$\pm$ 0.029} & 0.308 \footnotesize{$\pm$ 0.027} & 0.353 \footnotesize{$\pm$ 0.024}  \\
\midrule
Sum & 13.583 & 13.531 & 12.610 & 12.254 & 9.929 & 9.917 \\
\midrule
Div. & 0.686 & 0.498 & 0.728 & 0.596 & 0.816 & 0.846 \\
\bottomrule
\end{tabular}
}
    \label{tab:pmo_full2}
\end{table}

\begin{figure}[ht!]
    \centering
    \includegraphics[width=0.98\linewidth]{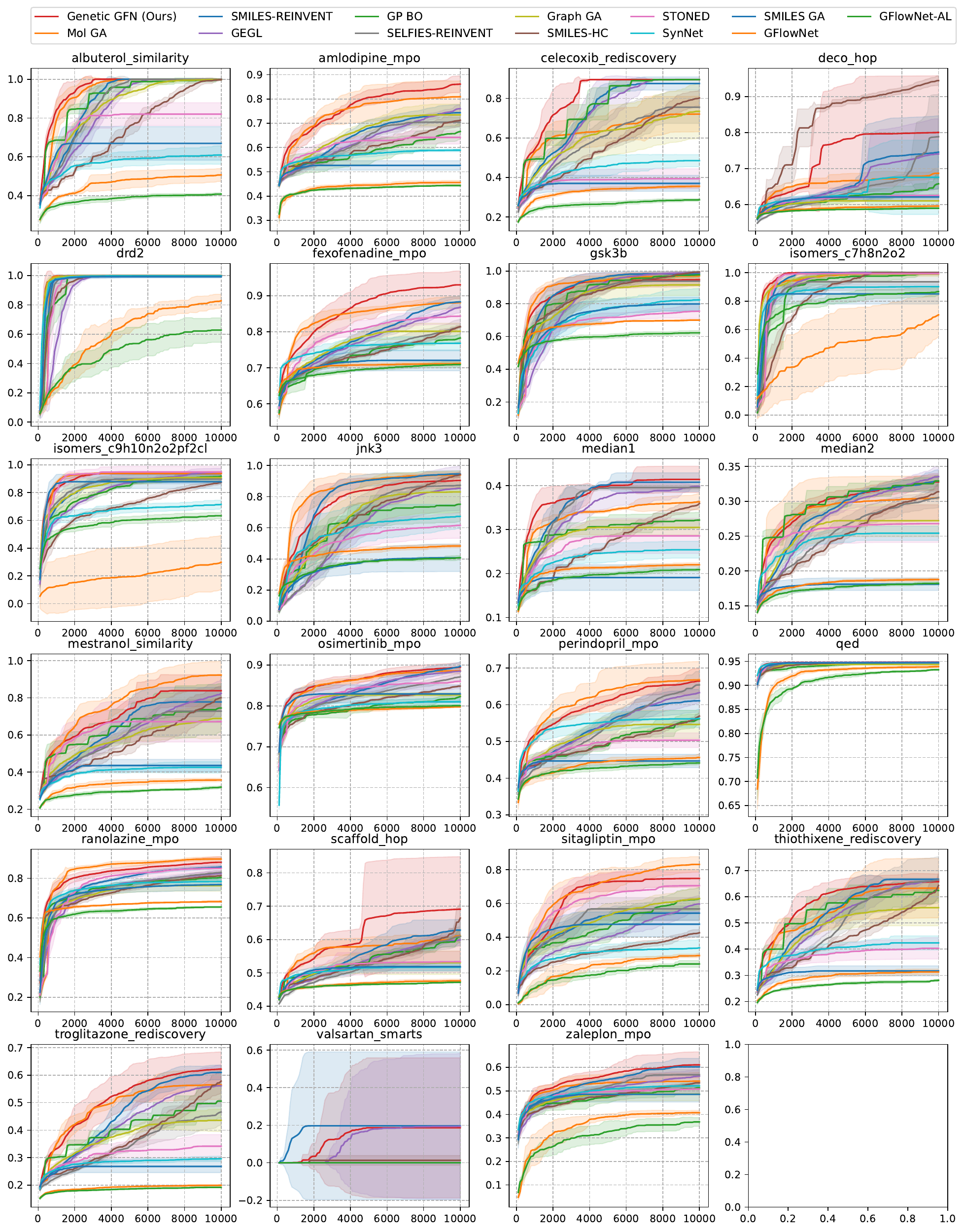}
    \caption{The optimization curves for 23 oracle score functions.}
    \label{fig:pmo_full}
\end{figure}

\clearpage
\subsubsection{Ranks with various metrics}

According to the PMO benchmark, we also provide the rank of each method with various metrics. The results in \cref{tab:rank} show that \ours{} achieves the first place in total, not only in the AUC Top-10.

\begin{table}[ht!]
    \centering
    \caption{The ranks of 10 methods based on various performance metrics}
    \resizebox{0.95\linewidth}{!}{\begin{tabular}{llccccccc}
\toprule
 & &\multicolumn{3}{c}{AUC} & \multicolumn{3}{c}{Average score} & \multirow{2}{*}{Mean} \\
\cmidrule(r){3-5} \cmidrule(r){6-8}
Method & Category &Top-1 & Top-10 & Top-100 & Top-1 & Top-10 & Top-100 &  \\
\midrule
\ours{} (Ours)& Off-policy & \textbf{1} & \textbf{1} & \textbf{1} & \textbf{1} & \textbf{1} & \textbf{1} & \textbf{1.00} \\
Mol GA \cite{tripp2023genetic} & GA  & 2 & 2 & 2 & 5 & 3 & 2 & 2.67 \\
SMILES-REINVENT \cite{olivecrona2017molecular} & On-policy & 3 & 3 & 3 & 4 & 2 & 3 & 3.00 \\
GEGL \cite{ahn2020guiding} & Off-policy & 4 & 4 & 6 & 3 & 4 & 4 & 4.17 \\
SELFIES-REINVENT \cite{olivecrona2017molecular} & On-policy & 7 & 6 & 4 & 6 & 6 & 5 & 5.67 \\
GP BO \cite{tripp2021fresh} & Active Learning  & 6 & 5 & 5 & 7 & 7 & 6 & 6.00 \\
SMILES-LSTM-HC \cite{brown2019guacamol} & Off-policy & 5 & 8 & 10 & 2 & 5 & 7 & 6.17 \\
Graph GA \cite{jensen2019graph} & GA & 8 & 7 & 7 & 8 & 8 & 8 & 7.67 \\
STONED \cite{nigam2021stoned} & GA & 9 & 9 & 8 & 9 & 9 & 9 & 8.83 \\
SynNet \cite{gao2022synnet} & GA & 10 & 10 & 11 & 10 & 10 & 11 & 10.33 \\
SMILES GA \cite{brown2019guacamol} & GA & 11 & 11 & 9 & 11 & 11 & 10 & 10.50 \\
GFlowNet \cite{bengio2021flow}& Off-policy & 13 & 13 & 12 & 12 & 12 & 12 & 12.33 \\
GFlowNet-AL \cite{jain2022biological}& Active Learning & 12 & 12 & 13 & 13 & 13 & 13 & 12.67 \\
\bottomrule
\end{tabular} 
}
    \label{tab:rank}
\end{table}

\subsection{Further studies on \texttt{valsartan\_smarts (\#22)} }

Notably, we have observed that only a few methods achieve non-zero scores on \texttt{valsartan\_smarts (\#22)} in \cref{tab:auc10}.
The valsartan SMARTS targets molecules containing a SMARTS pattern related to valsartan while being characterized by physicochemical properties corresponding to the sitagliptin molecule \citep{meyers2021novo}. It measures the arithmetic means of several scores, including (1) binary score about whether it contains a certain SMARTS structure, (2) LogP, (3) TPSA, and (4) Bertz score. Since we utilize a TDC oracle function for evaluations, we provide our empirical observations here.

\begin{figure}[ht!]
    \centering
    \begin{subfigure}[b]{0.47\textwidth}
    \includegraphics[width=\textwidth]{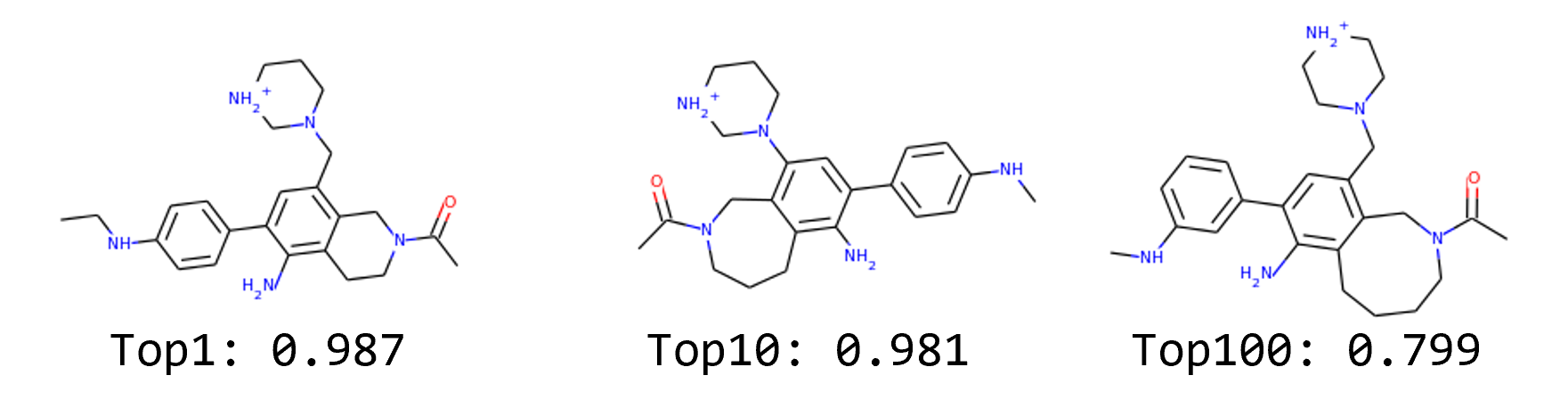}
    \caption{REINVENT (\texttt{seed 0})} 
    \end{subfigure}
    \hfill
    \begin{subfigure}[b]{0.47\textwidth}
    \includegraphics[width=\textwidth]{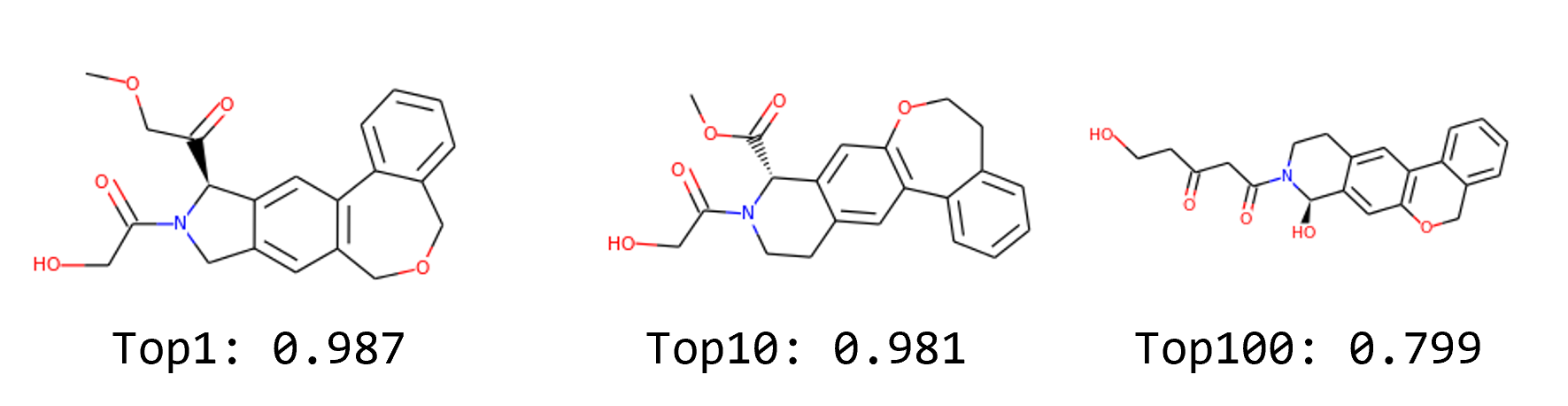}
    \caption{Genetic GFN (\texttt{seed 4})} 
    \end{subfigure}
    \caption{Examples of molecules with non-zero scores on \texttt{valsartan\_smarts (\#22)}} \label{fig:valsartan}
\end{figure}

Due to the binary score (1 if the certain SMARTS pattern is included), many tries terminate with 0. Especially with a limited number of oracle calls, generating molecules containing a certain sub-structure is notoriously hard. Other literature shows that other methods achieve high scores with more oracle calls \citep{hu2024molrl}. With 10K calls, even REINVENT and Genetic only succeed in finding non-zero score molecules once out of five independent runs.
Another observation is that methods (REINVENT, Genetic GFN, and GEGL) achieving non-zero scores all generate SMILES with RNN-based models. Thus, we have a conjecture that SMILES generation is effective in generating a certain SMARTS pattern.
We provide examples of generated molecules with non-zero \texttt{valsartan\_smarts} scores. Note that the other four seeds failed. Each run generates similar molecules (see Top1,10,100 samples in \cref{fig:valsartan} in the additional material), but the samples between the two runs (REINVENT and Genetic GFN) have different structures (the molecule distance between Top1 samples is 0.854).

\clearpage
\subsection{Additional results for ablation studies} \label{appnd:full_abl}

\subsubsection{Statistical analysis}
We provided the results of the t-tests of ablation studies. As shown in \cref{tab:t-test_abl}, the p-values for ablating genetic search are less than 0.05, so our genetic search is a statistically significant component.

\begin{table}[ht]
    \centering
    \caption{The results of t-tests of ablation studies.}
    \resizebox{0.98\linewidth}{!}{
\begin{tabular}{lrrrrrr}
\toprule
 & \multicolumn{4}{c}{Genetic Search} & \multicolumn{2}{c}{KL-divergence penalty} \\
\cmidrule(r){2-5} \cmidrule(r){6-7}
 & \ours & - $\{\text{GS}\}$ & \ours & - $\{\text{GS}\}$ + $\{\eps\text{-greedy} \}$ & \ours & - $\{\text{KL}\}$ \\
\midrule
Mean & 16.213 & 15.738 & 16.213 & 15.627 & 16.213 & 15.928 \\
Variance & 0.038 & 0.094 & 0.038 & 0.008 & 0.038 & 0.227 \\
Observations & 5 & 5 & 5 & 5 & 5 & 5 \\
Hypothesized Mean Diff. & 0 & & 0 & & 0 & \\
t Stat & 2.931 & & 6.109 & & 1.236 & \\
P($T \leq t$) one-tail & \textbf{0.011} & & \textbf{0.000} & & \textbf{0.136} \\
t Critical one-tail & 1.895 & & 1.943 & & 2.015 \\
P($T\leq t$) two-tail & \textbf{0.021} & & \textbf{0.001} & & \textbf{0.271} \\
t Critical two-tail & 2.365 & & 2.447 & & 2.571 \\
\bottomrule
\end{tabular} }
    \label{tab:t-test_abl}
\end{table}

\subsection{Results for diversity and synthesizability} \label{appnd:div_sa} 

We report the diversity and synthetic accessibility (SA) score of Top-100 molecules on each oracle.

\begin{figure}[ht!]
    \centering
    \includegraphics[width=0.7\linewidth]{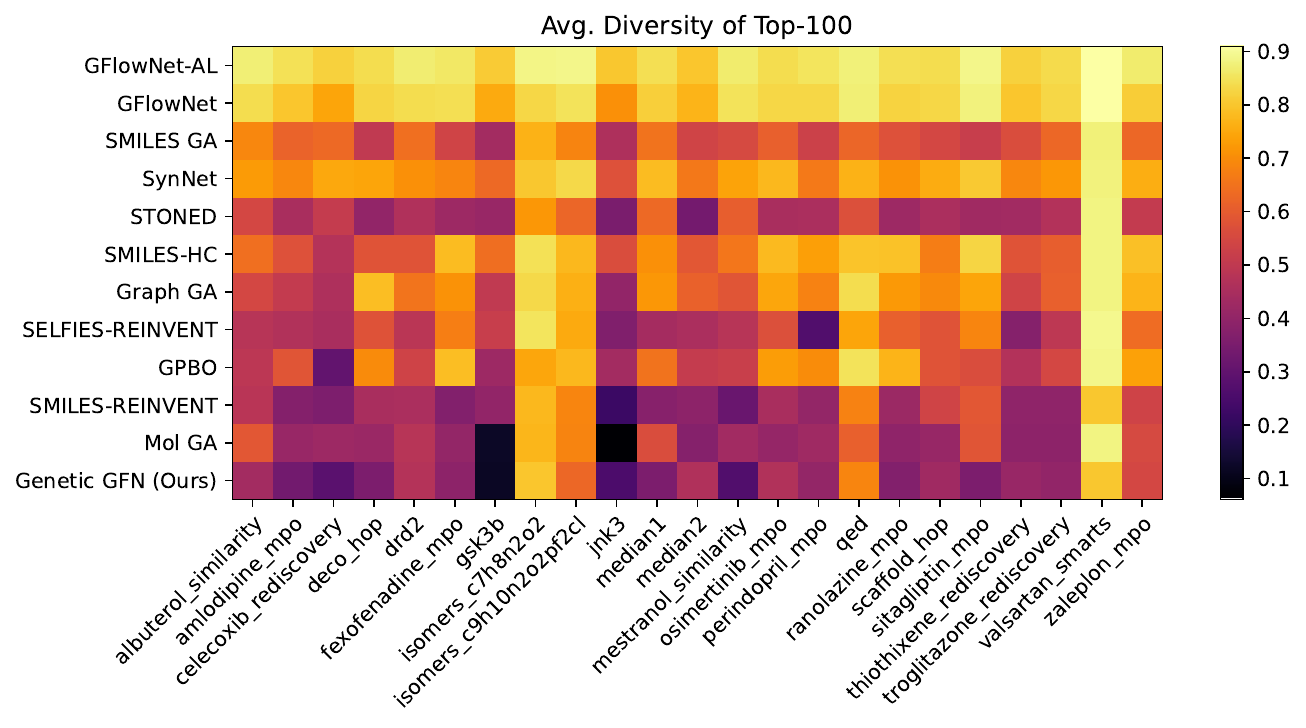}
    \caption{The average diversity of Top-100 molecules ($\uparrow$)}
    \label{fig:div}
\end{figure}

\begin{figure}[ht!]
    \centering
    \includegraphics[width=0.7\linewidth]{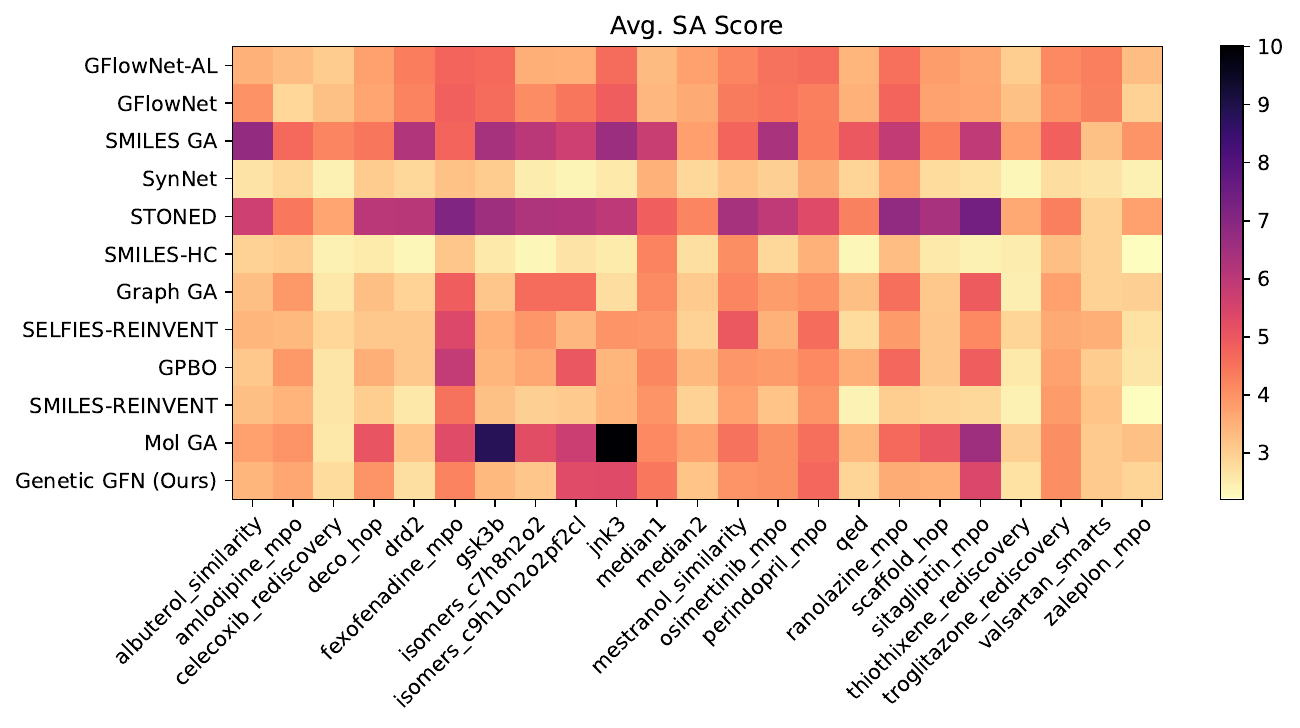}
    \caption{The average SA score of Top-100 molecules ($\downarrow$)}
    \label{fig:sa}
\end{figure}

\clearpage
\subsection{\ours{} with SELFIES generation} \label{appnd:selfies}

As our method employs a string-based sequence generation model, it is applicable to SELFIES representation. Following the same procedure and setup described in \cref{sec:method} and \cref{sec:exp_pmo}. The results in \cref{tab:selfies} demonstrate that \ours{} significantly outperforms SELFIES-REINVENT by achieving 14.986 compared to 14.152, not only the other high-ranked method, GP BO (14.264). Notably, our AUC Top-10 is higher than that of SELFIES-REINVENT in 19 oracles.

\begin{table}[ht]
    \centering
    \caption{Performance of \ours{} with SELFIES generations.} \label{tab:selfies}
    \resizebox{0.57\linewidth}{!}{\begin{tabular}{l|cc}
\toprule
Oracle & \makecell{SELFIES\\REINVENT} & \makecell{SELFIES\\ \ours{}} \\
\midrule
albuterol\_similarity & 0.867 $\pm$ 0.025 & 0.918 $\pm$ 0.031 \\
amlodipine\_mpo & 0.621 $\pm$ 0.015 & 0.711 $\pm$ 0.029 \\
celecoxib\_rediscovery & 0.588 $\pm$ 0.062 & 0.578 $\pm$ 0.049 \\
deco\_hop & 0.638 $\pm$ 0.016 & 0.631 $\pm$ 0.020 \\
drd2 & 0.953 $\pm$ 0.009 & 0.971 $\pm$ 0.005 \\
fexofenadine\_mpo & 0.740 $\pm$ 0.012 & 0.797 $\pm$ 0.012 \\
gsk3b & 0.821 $\pm$ 0.041 & 0.900 $\pm$ 0.042 \\
isomers\_c7h8n2o2 & 0.873 $\pm$ 0.041 & 0.952 $\pm$ 0.017 \\
isomers\_c9h10n2o2pf2cl & 0.844 $\pm$ 0.016 & 0.879 $\pm$ 0.031 \\
jnk3 & 0.624 $\pm$ 0.048 & 0.675 $\pm$ 0.140 \\
median1 & 0.353 $\pm$ 0.006 & 0.351 $\pm$ 0.034 \\
median2 & 0.252 $\pm$ 0.010 & 0.263 $\pm$ 0.014 \\
mestranol\_similarity & 0.589 $\pm$ 0.040 & 0.680 $\pm$ 0.076 \\
osimertinib\_mpo & 0.819 $\pm$ 0.005 & 0.849 $\pm$ 0.008 \\
perindopril\_mpo & 0.533 $\pm$ 0.024 & 0.551 $\pm$ 0.015 \\
qed & 0.940 $\pm$ 0.000 & 0.942 $\pm$ 0.000 \\
ranolazine\_mpo & 0.736 $\pm$ 0.008 & 0.785 $\pm$ 0.013 \\
scaffold\_hop & 0.521 $\pm$ 0.014 & 0.531 $\pm$ 0.020 \\
sitagliptin\_mpo & 0.492 $\pm$ 0.055 & 0.590 $\pm$ 0.018 \\
thiothixene\_rediscovery & 0.497 $\pm$ 0.043 & 0.527 $\pm$ 0.036 \\
troglitazone\_rediscovery & 0.342 $\pm$ 0.022 & 0.387 $\pm$ 0.087 \\
valsartan\_smarts & 0.000 $\pm$ 0.000 & 0.000 $\pm$ 0.000 \\
zaleplon\_mpo & 0.509 $\pm$ 0.009 & 0.518 $\pm$ 0.016 \\
\midrule
Sum & 14.152 & 14.986 \\
Diversity & 0.555 & 0.528 \\
\bottomrule
\end{tabular}
}
\end{table}

\subsection{\ours{} with string-based genetic search}

We also additionally provide experiments that incorporate STONED (GA with SELFIES)\citep{nigam2021stoned} as an exploration strategy to guide GFN training instead of Graph GA. Note that STONED only utilizes mutations since designing valid crossover with string representation is challenging.

\begin{table}[ht]
    \centering
    \caption{Results with different genetic search algorithms}
    \label{tab:my_label}
    \resizebox{0.5\linewidth}{!}{
    \begin{tabular}{lcc}
    \toprule
     & \ours{} & \makecell{\ours{}\\with STONED}  \\
    \midrule
    AUC Top-1
    & 16.527 \footnotesize{$\pm$ 0.043}
    & 15.806 \footnotesize{$\pm$ 0.037} \\
    AUC Top-10 
    & 16.213 \footnotesize{$\pm$ 0.042}
    & 15.439 \footnotesize{$\pm$ 0.037} \\
    AUC Top-100 
    & 15.516 \footnotesize{$\pm$ 0.041}
    & 14.870 \footnotesize{$\pm$ 0.036} \\
    \bottomrule
    \end{tabular}
    }
\end{table}

\clearpage
\subsection{Experiments with varying hyperparameters} \label{appnd:sensistivity}

In this subsection, we verify the robustness of \ours{} for differing hyperparameter setups. We conduct experiments by varying the number of GA generation (refining loops), offspring size, and the number of training inner loops. The results show that our results are robust to each hyperparameter setup by achieving similar or better performance compared to Mol GA in all tested configurations.

\begin{table}[ht!]
    \centering
    \caption{Ablation studies for the number of GA generation. `$\times 0$' stands for the results of TB loss training without GA explorations.}
    \resizebox{0.45\linewidth}{!}{\begin{tabular}{l|cccc}
\toprule
Oracle &  $\times 0$  & $\times1$ & $\times 2$ & $\times 3$   \\

\midrule
\textbf{AUC Top-1} & 16.070 & 15.968 & \textbf{16.527} & 16.040 \\
\textbf{AUC Top-10} & 15.738 & 15.615 & \textbf{16.213} & 15.735 \\
\textbf{AUC Top-100} & 15.030 & 14.909 & \textbf{15.516} & 15.074 \\
\textbf{Diversity} & \textbf{0.479} & 0.470 & 0.432 & 0.440 \\
\bottomrule
\end{tabular} 
}
    \label{tab:generation}
\end{table}

\begin{table}[ht!]
    \centering
    \caption{Results by varying the offspring size. }
    \resizebox{0.45\linewidth}{!}{\begin{tabular}{l|cccc}
\toprule
Oracle &  4  & 8  & 16 & 32   \\

\midrule
\textbf{AUC Top-1} & 16.174 & \textbf{16.527} & 15.984 & 15.963 \\
\textbf{AUC Top-10} & 15.846 & \textbf{16.213} & 15.669 & 15.621 \\
\textbf{AUC Top-100} & 15.175 & \textbf{15.516} & 14.977 & 14.858 \\
\textbf{Diversity} & \textbf{0.458} & 0.432 & 0.452 & 0.437 \\
\bottomrule
\end{tabular} 
}
    \label{tab:offspring}
\end{table}

\begin{table}[ht!]
    \centering
    \caption{Results by varying the number of training inner loops.}
    \resizebox{0.4\linewidth}{!}{\begin{tabular}{l|ccc}
\toprule
Oracle &  $\times 4$ & $\times 8$  & $\times 16$   \\

\midrule
\textbf{AUC Top-1} & 16.125 & \textbf{16.527} & 16.049  \\
\textbf{AUC Top-10} & 15.768 & \textbf{16.213} & 15.717 \\
\textbf{AUC Top-100} & 15.175 & \textbf{15.516} & 14.977 \\
\textbf{Diversity} & 0.423 & 0.432 & \textbf{0.511} \\
\bottomrule
\end{tabular} 
}
    \label{tab:training_loop}
\end{table}


\newpage
\section*{NeurIPS Paper Checklist}

\begin{enumerate}

\item {\bf Claims}
    \item[] Question: Do the main claims made in the abstract and introduction accurately reflect the paper's contributions and scope?
    \item[] Answer: \answerYes{} 
    \item[] Justification: The abstract and introduction clearly state our main research claim, which is consistent with the experimental results. 

\item {\bf Limitations}
    \item[] Question: Does the paper discuss the limitations of the work performed by the authors?
    \item[] Answer: \answerYes{} 
    \item[] Justification: We include our limitations and future works in the discussion section.

\item {\bf Theory Assumptions and Proofs}
    \item[] Question: For each theoretical result, does the paper provide the full set of assumptions and a complete (and correct) proof?
    \item[] Answer: \answerNA{} 
    \item[] Justification: The paper does not include theoretical results. 

    \item {\bf Experimental Result Reproducibility}
    \item[] Question: Does the paper fully disclose all the information needed to reproduce the main experimental results of the paper to the extent that it affects the main claims and/or conclusions of the paper (regardless of whether the code and data are provided or not)?
    \item[] Answer: \answerYes{} 
    \item[] Justification: We provide implementation details, including pseudo-codes.

\item {\bf Open access to data and code}
    \item[] Question: Does the paper provide open access to the data and code, with sufficient instructions to faithfully reproduce the main experimental results, as described in supplemental material?
    \item[] Answer: \answerYes{} 
    \item[] Justification: We provide the anonymized link of our implementation. Also, the paper uses the public dataset and benchmark; we provide accessible links in the supplemental material.

\item {\bf Experimental Setting/Details}
    \item[] Question: Does the paper specify all the training and test details (e.g., data splits, hyperparameters, how they were chosen, type of optimizer, etc.) necessary to understand the results?
    \item[] Answer: \answerYes{} 
    \item[] Justification: We provide all details regarding experiments in the appendix and the anonymized codes.

\item {\bf Experiment Statistical Significance}
    \item[] Question: Does the paper report error bars suitably and correctly defined or other appropriate information about the statistical significance of the experiments?
    \item[] Answer: \answerYes{} 
    \item[] Justification: The experimental results include standard deviation with independent multiple runs. In addition, statistical tests are conducted for the main results and self-ablation studies.

\item {\bf Experiments Compute Resources}
    \item[] Question: For each experiment, does the paper provide sufficient information on the computer resources (type of compute workers, memory, time of execution) needed to reproduce the experiments?
    \item[] Answer: \answerYes{} 
    \item[] Justification: We provide the computer resources in the appendix (due to lack of spaces).
    
\item {\bf Code Of Ethics}
    \item[] Question: Does the research conducted in the paper conform, in every respect, with the NeurIPS Code of Ethics \url{https://neurips.cc/public/EthicsGuidelines}?
    \item[] Answer: \answerYes{} 
    \item[] Justification: We have checked the paper according to the NeurIPS Code of Ethics; our study does not include human participation and privacy-related data. 

\item {\bf Broader Impacts}
    \item[] Question: Does the paper discuss both potential positive societal impacts and negative societal impacts of the work performed?
    \item[] Answer: \answerYes{} 
    \item[] Justification: We discuss the broader impact of this study at the end of the paper.

\item {\bf Safeguards}
    \item[] Question: Does the paper describe safeguards that have been put in place for responsible release of data or models that have a high risk for misuse (e.g., pretrained language models, image generators, or scraped datasets)?
    \item[] Answer: \answerNA{} 
    \item[] Justification: Our study is conducted using the public dataset, and all experimental tasks are from previous publications.

\item {\bf Licenses for existing assets}
    \item[] Question: Are the creators or original owners of assets (e.g., code, data, models), used in the paper, properly credited and are the license and terms of use explicitly mentioned and properly respected?
    \item[] Answer: \answerYes{} 
    \item[] Justification: The paper clearly states the source and license of the original code, data, and models in the appedix. 

\item {\bf New Assets}
    \item[] Question: Are new assets introduced in the paper well documented and is the documentation provided alongside the assets?
    \item[] Answer: \answerNA{} 
    \item[] Justification: Our experiments are conducted using the already published datasets and benchmarks.

\item {\bf Crowdsourcing and Research with Human Subjects}
    \item[] Question: For crowdsourcing experiments and research with human subjects, does the paper include the full text of instructions given to participants and screenshots, if applicable, as well as details about compensation (if any)? 
    \item[] Answer: \answerNA{} 
    \item[] Justification: Our study does not involve crowdsourcing nor human subjects.

\item {\bf Institutional Review Board (IRB) Approvals or Equivalent for Research with Human Subjects}
    \item[] Question: Does the paper describe potential risks incurred by study participants, whether such risks were disclosed to the subjects, and whether Institutional Review Board (IRB) approvals (or an equivalent approval/review based on the requirements of your country or institution) were obtained?
    \item[] Answer: \answerNA{} 
    \item[] Justification: Our study does not involve crowdsourcing nor human subjects.

\end{enumerate}

\end{document}